\documentclass[a4paper,fleqn,usenatbib]{mnras}
\usepackage{graphicx}
\usepackage{amsmath}
\usepackage{mathptmx}
\usepackage{txfonts}
\usepackage[T1]{fontenc}
\usepackage{ae,aecompl}
\usepackage{upgreek}
\usepackage{aas_macros}
\usepackage{lmodern}
\title[The clustering of submillimetre galaxies]{The SCUBA-2 Cosmology Legacy Survey: The clustering of submillimetre galaxies in the UKIDSS UDS field}
\author[A. Wilkinson et. al.]
{Aaron Wilkinson$^{1}$\thanks{E-mail: ppxakw@nottingham.ac.uk (AW)},
Omar Almaini$^{1}$, Chian-Chou Chen$^{2,3}$, Ian Smail$^{2,3}$,
\newauthor{Vinodiran Arumugam$^{4,5}$, Andrew Blain$^6$, Edward L. Chapin$^{7}$, Scott C. Chapman$^{8}$,}
\newauthor{Christopher J. Conselice$^{1}$, William I. Cowley$^{3}$, James S. Dunlop$^{4}$, Duncan Farrah$^{9}$}
\newauthor{James Geach$^{10}$, William G. Hartley$^{11}$, Rob J. Ivison$^{4,5}$, David T. Maltby$^{1}$,}
\newauthor{Micha\l{} J. Micha\l{}owski$^{4}$, Alice Mortlock$^{4}$, Douglas Scott$^{12}$, Chris Simpson$^{13}$,}
\newauthor{James M. Simpson$^{4}$, Paul van der Werf$^{14}$, Vivienne Wild$^{15}$} \\
$^{1}$School of Department of Physics and Astronomy, University of Nottingham, University Park, Nottingham, NG7 2RD, UK\\
$^{2}$Centre for Extragalactic Astronomy, Department of Physics, Durham University, South Road, Durham, DH1 3LE, UK\\
$^{3}$Institute for Computational Cosmology, Durham University, South Road, Durham, DH1 3LE, UK\\
$^{4}$SUPA$\dagger$, Institute for Astronomy, University of Edinburgh, Royal Observatory, Blackford Hill, Edinburgh, EH9 3HJ, UK\\
$^{5}$European Southern Observatory, Karl Schwarzschild Strasse 2, Garching, Germany\\
$^6$Department of Physics $\&$ Astronomy, University of Leicester, University Road, Leicester, LE1 7RH, UK\\
$^{7}$Herzberg Astronomy and Astrophysics, National Research Council Canada, 5071 West Saanich Road, Victoria, BC V9E 2E7, Canada\\
$^{8}$Department of Physics and Atmospheric Science, Dalhousie University, Halifax, NS B3H 3J5, Canada\\
$^{9}$Department of Physics, Virginia Tech, Blacksburg, VA 24061, USA\\
$^{10}$Centre for Astrophysics Research, University of Hertfordshire, College Lane, Hatfield AL10 9AB, UK\\
$^{11}$ETH Z{\"u}rich, Institut f{\"u}r Astronomie, HIT J 11.3, Wolfgang-Pauli-Strasse 27, Z{\"u}rich, CH-8093, Switzerland\\
$^{12}$Department of Physics $\&$ Astronomy, University of British Columbia, 6224 Agricultural Road, Vancouver, BC V6T 1Z1, Canada\\
$^{13}$Astrophysics Research Institute, Liverpool John Moores University, Liverpool Science Park, 146 Brownlow Hill, Liverpool, L3 5RF, UK\\
$^{14}$Leiden Observatory, Leiden University, P.O. Box 9513, NL-2300 RA Leiden, The Netherlands \\
$^{15}$SUPA\thanks{Scottish Universities Physics Alliance}, School of Physics and Astronomy, University of St Andrews, North Haugh, St Andrews, KY16 9SS, U.K.
}
\begin{document}

\date{Accepted yyyy mm dd. Received yyyy mm dd; in original form yyyy mm dd}

\pagerange{\pageref{firstpage}--\pageref{lastpage}} \pubyear{0000}

\maketitle

\label{firstpage}
%\vspace{-0.8cm}
\begin{abstract}
Submillimetre galaxies (SMGs) are among the most luminous dusty galaxies in the Universe, 
but their true nature remains unclear; are SMGs the
progenitors of the massive elliptical galaxies we see in the local Universe, or are they just a short-lived phase among more typical star-forming galaxies?
To explore this problem further, we investigate the clustering of  
SMGs identified in the SCUBA-2 Cosmology Legacy Survey. We use a catalogue of submillimetre ($850\,\umu$m) source identifications derived using a combination of radio counterparts  
and colour/IR selection to analyse a sample of 610 SMG counterparts in the UKIDSS Ultra Deep Survey (UDS), making this the largest high redshift 
sample of these galaxies to date.
 Using angular cross-correlation techniques, we estimate the halo masses for this large sample of SMGs and 
compare them with passive and star-forming galaxies selected in the same field.
We find that SMGs, on average, occupy high-mass dark matter halos (M$_{\text{halo}} >10^{13}$\,M$_{\odot}$) at redshifts
$z > 2.5$, consistent with being the progenitors of massive quiescent galaxies in present-day galaxy clusters. We also find evidence of downsizing, in which SMG activity
shifts to lower mass halos at lower redshifts. In terms of their clustering and halo masses, SMGs appear to be consistent with other star-forming
galaxies at a given redshift.

\end{abstract}


%%%%%%%%%%%%%%%%%%%%%%%%%%%%%%%%%%%%%%%%%%%%%%%%%%%%%%%%%%%%%%%%%%%%%%%%%%%%%%%%%%%%%%%%%%%%%%%%%%%%%%%%%%%%%%%%%%%%%%%%%%%%%%%%%%%%%%%%%%%%%%%%%%%%%%%%%%%%%%%%
%\vspace{-0.6cm}
\section{Introduction}
\label{sec:intro}
One key discovery in astronomy to date is the 
bimodality in the galaxy population, whose origin is still greatly debated.
 The galaxy population is split into two distinct types both locally and at high redshift: passively-evolving
 red sequence galaxies and the star-forming blue 
cloud galaxies \citep[e.g.,][]{2001AJ....122.1861S, 2004AIPC..743..106B, 2004ApJ...608..752B, 2009ApJ...706L.173B}. 
The quenching of star formation is believed to cause blue cloud galaxies to migrate to the red sequence. A difference in morphology 
is also observed. A large fraction of passive galaxies have spheroidal 
early-type morphologies, in contrast to the star-forming galaxies, 
which tend to form disk-like structures \citep{1998ARA&A..36..189K}.

Various galaxy evolution models have been proposed to explain the morphological transformation and the quenching of star formation in disk galaxies
\cite[e.g.][]{2005Natur.433..604D, 2006MNRAS.367..864C, 2006MNRAS.368....2D, 2006ApJS..163....1H, 2008MNRAS.391..481S, 2009ApJ...707..250M, 2016arXiv160107907T}.
Many of these models invoke major merger events, originally proposed by
\citet{1977egsp.conf..401T}. A merging of two or more galaxies can result in a starburst phase in 
which the merged galaxy experiences a short-lived burst of compact star formation. In the aftermath,
 stellar or active galactic nuclei (AGN) feedback can rapidly expel remaining gas from the
galaxy \citep[e.g.,][]{1998A&A...331L...1S,2005ApJ...630..705H, 2016arXiv160107907T}. Alternative models produce compact spheroids using inflow of cold gas,
which lead to disk instabilities and contraction \citep[e.g.,][]{2009ApJ...703..785D}. For a galaxy to remain 
quenched in its star formation, AGN feedback is required to keep any gas sufficiently 
heated \citep{2006MNRAS.368L..67B}.

A merger-induced starburst may be responsible for the formation of the most massive (M$_{*}>10^{11}$M$_{\odot}$) elliptical galaxies in the local Universe.
Evidence tentatively suggests that these galaxies were assembled at high redshifts ($z\sim3$--$5$), with the preceding starburst event taking place on short timescales
of $\sim500\,$Myr \citep{2010MNRAS.404.1775T}. One way to link galaxy populations formed at different redshifts is to derive their halo masses. 
Galaxy clustering provides a powerful method to %to determine the projected progenitors and descendants of various galaxy populations, 
constrain halo masses, particularly at high redshifts.
Various clustering studies so far have revealed that passive galaxies cluster more strongly than 
their star-forming counterparts \citep{2002MNRAS.332..827N, 2009MNRAS.399..878R, 2009ApJ...691.1879W, 2010MNRAS.407.1212H, 2013MNRAS.431.3045H}
 and preferentially reside in more massive dark matter halos. \cite{2013MNRAS.431.3045H} analysed the clustering of passive galaxies 
by calculating 2-point angular correlation functions for photometrically selected samples to $z\sim3$, splitting their samples into 
bins of redshift and stellar mass. They found that passive galaxies are the most 
strongly clustered, residing in halos of mass M$_{\text{halo}}>5\times10^{12}$M$_{\odot}$.
By establishing the typical host halo masses of high redshift galaxies and the evolution
of these halo masses to the present-day Universe, we can identify the possible progenitors of local massive elliptical galaxies.

A rare and interesting class of high redshift galaxies is the population of ultra-luminous dusty galaxies with bright flux densities in the submillimetre waveband
\citep{1997ApJ...490L...5S, 1998Natur.394..248B, 1998Natur.394..241H}.
Known as submillimetre galaxies (SMGs), they appear to have redshift distributions peaking at $z\sim2.5$ \citep[e.g. ][]{2005ApJ...622..772C,2014ApJ...788..125S},
occupying the same epoch associated with
the peak activity of luminous AGN activity \citep{2006AJ....131.2766R, 2011ApJ...728...56A}. The
extreme luminosities observed in these dusty sources are thought to be powered by intense short-lived ($\sim100$\,Myrs) starbursts
\citep{2005ApJ...632..736A, 2006ApJ...640..228T, 2008ApJ...680..246T, 2011MNRAS.412.1913I}. 
Many previous studies \citep[e.g.,][]{1998Natur.394..241H, 1999ApJ...515..518E, 2006MNRAS.371..465S, 2011MNRAS.412..295T} suggested that SMGs may be the progenitors
of the most massive elliptical galaxies we see in the local Universe today.
This scenario is tentatively supported by numerous clustering studies of SMGs identified in the long submillimetre wavelength bands
\citep[$850$--$1100\,\umu$m;][]{2003ApJ...582....6W, 2004ApJ...611..725B, 2009ApJ...707.1201W, 2011ApJ...733...92W, 2012MNRAS.421..284H},
%blain, webb = 850; weiss, hickox = 870, williams = 1100
where observations of a strong clustering amplitude suggested SMGs
resided in high-mass ($10^{12}$--$10^{13}$h$^{-1}$Mpc) dark matter halos. Clustering studies of SMGs detected in the Herschel field
\citep[with shorter submillimetre wavelengths $250$--$500\,\umu$m; e.g.,][]{Cooray+2010,2010A&A...518L..11M,2012ApJ...753...23M,2012MNRAS.426.3455V}
also confirmed strong SMG clustering signals. In fact, there may be evidence
of an evolution of clustering with redshift, with \citet{2010A&A...518L..11M} and \citet{2012MNRAS.426.3455V} reporting low clustering strengths for SMGs in redshifts $z<0.3$.
However, these studies may
have been selecting different galaxy populations at low redshifts, compared to those identified at higher redshifts.
In addition, many of these previous studies analysed only modest samples of SMGs (at the most $\sim100$), and consequently halo masses were
difficult to constrain.

Recently however, we have obtained a much larger sample of SMGs. Technological advances with bolometer cameras, such as the SCUBA-2 camera on
the 15\,m James Clark Maxwell Telescope (JCMT), have allowed us to undertake submillimetre surveys over square degree areas down to mJy sensitivity limits
\citep[such as the SCUBA-2 Cosmology Legacy Survey or S2CLS,][]{2013MNRAS.430.2513H, 2013MNRAS.432...53G}.
\citet{2016arXiv160102630C} used the radio and optical/IR data from the UDS to
 identify $\sim 1000$ SMGs in the S2CLS field, making this the largest SMG sample so far in the $850\,\umu$m waveband.
The increase of the sample size allowed \citet{2016arXiv160102630C} to make a clear detection in the 2-point angular correlation function.
However, the measurements still suffer from large uncertainties, and the evolution of SMG clustering is not constrained at all.

In this work, we make use of a cross-correlation 
technique to statistically associate the sample of SMGs to a much larger K-band selected galaxy sample,
which allows us to infer the dark matter halo mass with much greater confidence, 
and to constrain the evolution of SMG clustering for the first time.
Fundamentally, by studying the dark matter halos inhabited by these rare galaxies,
we can identify their progenitors and descendants,
helping us to understand the evolutionary link between extreme star-forming galaxies
and those on the red sequence.

The structure of this paper is as follows: Section 2 contains the discussion of our 
data sets and sample selections; Section 3 describes our clustering analysis in greater 
detail; in Section 4 we show the results and discuss the implications; we end with our 
conclusions and further work in Sections 5 and 6. Throughout 
this paper we assume a $\lambda$-CDM cosmology with 
$\Omega_{\text{M}}=0.3$, $\Omega_{\lambda}=0.7$, H$_0=70$\,kms$^{-1}$Mpc$^{-1}$ and $\sigma_8=0.9$. All magnitudes 
are given in the AB system, unless otherwise stated.
\vspace{-0.3cm}
%%%%%%%%%%%%%%%%%%%%%%%%%%%%%%%%%%%%%%%%%%%%%%%%%%%%%%%%%%%%%%%%%%%%%%%%%%%%%%%%%%%%%%%%%%%%%%%%%%%%%%%%%%%%%%%%%%%%%%%%%%%%%%%%%%%%%%%%%%%%%%%%%%%%%%%%%%%%%%%%

\section{UDS DATA SET AND SAMPLE SELECTION}
\label{sec:data}

In this section, we introduce the SMG sample obtained from the SCUBA-2 Cosmology Legacy Survey (S2CLS) map
of the UKIRT Infrared Deep Sky Survey (UKIDSS), Ultra-Deep Survey (UDS) field, as well as passive and normal star-forming galaxies selected from the latter survey.
We use $K$-band selected samples from the UKIDSS UDS Data Release 8, complemented by matching multi-wavelength 
photometric data. Covering 0.77 square degrees, the UDS is a deep survey in the $J, H$
and $K$ wavebands. Reaching a depth of $K=24.6$, it is the deepest near-infrared survey 
to date over such a large area. The final UDS data release (planned for mid 2016) will achieve estimated depths of $J=25.4, H=24.8$ and $K=25.3$.
The optical/IR catalogue used in this work is described in
\citet{2013MNRAS.431.3045H}.

The UDS field is also covered by data in the $B, V, R, i', z'$ optical bands from the Subaru XMM-Newton Deep Survey 
(SXDS, \citealt{2008ApJS..176....1F}), the $u$-band from the CHFT Megacam and three IR bands (two near-IR and one mid-IR)
from the \textit{Spitzer} UDS Legacy Program (SpUDS, PI:Dunlop).
SpUDS provides data in channels 1 and 2 of IRAC ($3.6$ and $4.5\,\umu$m, respectively)
as well as in the MIPS $24\,\umu$m waveband. After masking out bad 
regions and bright stars found in the UDS image, the co-incident area of these combined data sets is 0.62 square degrees.
 Finally, we use X-ray \citep{2008ApJS..179..124U} and radio 
\citep{2006MNRAS.372..741S} observations to remove luminous AGN.

The submillimetre galaxies were identified using the final $850\,\umu$m S2CLS maps of
the UDS field (\citealt{2016arXiv160102630C}; Geach et al. in preparation). Reaching a median depth of $\sim0.9$\,mJy per beam, these maps are taken from the SCUBA-2 camera
at the JCMT. The SCUBA-2 map in the UDS field has a noise of 0.82\,mJy per beam at the deepest part, with rms noise $<1.3$\,mJy
over $\sim1.0$ degree$^{2}$. Compared to the LABOCA survey in the ECDF-S (LESS; \citealt{2009ApJ...707.1201W}),
 our map is $\sim40\%$ deeper in sensitivity and has a $\sim30\%$ improvement in spatial resolution, producing
the largest sample of SMG identifications to date ($6\times$ larger than the LESS sample).
 We employ a robust sample of 716 SMGs detected at a significance of
$>4\sigma$ from \citet{2016arXiv160102630C}, in our clustering analysis.

\vspace{-0.3cm}
%%%%%%%%%%%%%%%%%%%%%%%%%%%%%%%%%%%%%%%%%%%%%%%%%%%%%%%%%%%%%%%%%%%%%%%%%%%%%%%%%%%%%%%%%%%%%%%%%%%%%%%%%%%%%%%%%%%%%%%%%%%%%%%%%%%%%%%%%%%%%%%%%%%%%%%%%%%%%%%%

\subsection{Photometric redshifts and stellar masses}
\label{sec:photored}

We use photometric redshifts in our analysis, obtained from the combination of deep 
photometry and a sample of over 3000 secure spectroscopic redshifts. 
Most redshifts at $z>1$ were obtained from the UDSz ESO Large Programme (ID: 180.A-0776; PI: Almaini).

Using the \textsc{eazy} template-fitting package \citep{2008ApJ...686.1503B}, photometric redshift
probability distributions were calculated for each object through a maximum likelihood analysis
(see \citealt{2013MNRAS.431.3045H} and \citealt{2013MNRAS.433.1185M} for further details).
The template fitting made use of the six standard \textsc{eazy} templates and an additional template, a combination of the bluest
\textsc{eazy} template and a small amount of Small Magellanic Cloud-like extinction \citep{1984A&A...132..389P}.
 We train the fitting on the spectroscopic sample
by comparing their spectroscopic redshifts to their derived photometric redshifts. Excluding outliers ($(z_{\rm{phot}}-z_{\rm{spec}})/(1+z)>0.15$, $<4\%$ of objects),
we find the dispersion in
$(z_{\rm{phot}}-z_{\rm{spec}})/(1+z)$ to be $\sigma=0.031$. We have 242 passive and 1,131 star-forming galaxies (separated using our $UVJ$ selection, as discussed in
Section 2.3) with spectroscopic redshifts, which are found to be in good agreement with their derived photometric redshifts. We find the
dispersions in $(z_{\rm{phot}}-z_{\rm{spec}})/(1+z)$ to be $\sigma=0.023$ and $\sigma=0.033$, respectively, with outlier fraction of $\sim3\%$ for both populations.
However, we must note here that we have only 10 SMGs for which we have spectroscopic data, stressing the importance of using
photometrically-derived redshifts in this analysis. Investigating the robustness of photometric redshifts for this population is difficult due to low number of sources. Nevertheless,
we estimate a dispersion of $\sigma=0.027$, with an outlier fraction of $10\%$ (one object),
 giving us confidence that photometric redshifts derived for SMGs are reasonable.
Of the 10 SMGs with spectroscopic redshifts, only 5 reside at $z>1$, the minimum redshift limit we consider for our analysis.

For our clustering analysis, we make use of the full probability distribution, similar to the approach adopted by \citet{2011ApJ...728...46W} and \citet{2013MNRAS.431.3045H}.
A galaxy can have multiple entries across each redshift slice,
whose contribution is weighted by the integrated probability between the limits of the redshift slices. We recompute stellar masses and rest-frame colours for each entry, at the
minimum $\chi^2$ redshift within the interval.
We note that the redshift probability distributions typically become broader with increasing redshift,
as also suggested from the overall comparison of spectroscopic and photometric redshifts, but this information is contained
within the weighting system; pairs of galaxies with broad redshift
distributions will make a smaller contribution to the clustering analysis.  Further spectroscopic redshifts for SMGs will be
required to determine if the probability distributions are accurate for this population. We note, however, that using the single
best-fit redshifts in our clustering analysis gives similar results, albeit with larger uncertainties.

Stellar masses were obtained from the same approach used by \citet{2013MNRAS.431.3045H} and \citet{2013MNRAS.433.1185M, 2015MNRAS.447....2M}. We fitted 
$u'BVRi'z'JHK$ bands and IRAC Channels 1 and 2 to a grid of spectral energy distributions (SEDs). These SEDs were composed from 
\citet{2003MNRAS.344.1000B} stellar population models and a Chabrier initial mass function was assumed for the calculations. The star formation histories (with a variety of
ages, dust extinctions and metallicities) were modelled by an exponentially declining star formation.
The grid of synthetic SEDs were scaled to the $K$-band magnitude of the galaxy we wished to fit. Each scaled template was then fitted to the observed photometry,
with stellar masses computed from the best-fitting template. We refer readers to the \citet{2013MNRAS.431.3045H} and \citet{2013MNRAS.433.1185M} papers for a more
in-depth discussion of the stellar mass derivation.

\vspace{-0.3cm}
%%%%%%%%%%%%%%%%%%%%%%%%%%%%%%%%%%%%%%%%%%%%%%%%%%%%%%%%%%%%%%%%%%%%%%%%%%%%%%%%%%%%%%%%%%%%%%%%%%%%%%%%%%%%%%%%%%%%%%%%%%%%%%%%%%%%%%%%%%%%%%%%%%%%%%%%%%%%%%%%

\subsection{Sample selection - submillimetre galaxies}
\label{sec:sampleselect}

\citet[hereafter C16]{2016arXiv160102630C} identified the counterpart candidates for the SCUBA-2 detected sources by using a combined radio and optical/infrared 
 colour method. This was
trained on a sample of 52 SMGs in the UDS region, with identifications provided by the Atacama Large Millimeter/submillimeter Array \citep[ALMA;][]{2015ApJ...799...81S, 2015ApJ...807..128S},
as well as the ALMA-identified ALESS survey \citep{2013ApJ...768...91H}.
 We briefly summarise the method here, referring readers to C16 for further details. C16 first identified any radio sources from a deep 1.46\,Hz Very Large Array (VLA) survey of the field
that matched the SMGs to within 8.7$"$ of the ALMA primary beam. 
Using the ALMA training sets in the UDS and ALESS, they showed that these radio-identified SMGs usually corresponded to the brightest component in a single dish source.
In addition, C16 also included the radio-undetected, $K$-selected sources that satisfied the Optical-IR Triple Colour (OIRTC) criteria.
C16 used a new colour selection technique that is based on the fact that SMGs occupy a distinct parameter space in $z-K$, $K-m_{3.6}$ and $m_{3.6}-m_{4.5}$ colours,
making it a useful method to separate SMGs from non-SMG field galaxies.
This OIRTC selection is suitable for identifying the fainter counterparts to the SCUBA-2 sources,
complementary to the radio detection of the brightest SMGs.

Consequently, we used the radio and OIRTC identifications from C16 to construct the sample of SMG counterparts used for this work. C16 adopted a search radius of 8.7$"$ within the SCUBA-2
beam, in which this radius corresponds to a $4$\,$\sigma$ positional uncertainty for a $4$\,$\sigma$ SCUBA-2 detection.
We note that the search radius is the same as the width of the ALMA primary
beam used to derive the training sample. Based on the ALMA training sample,
the combined radio$+$OIRTC selection method identified SMGs with an accuracy of $83^{+17}_{-19}\%$ and a completeness of $67\pm14\%$. Here we define the accuracy to be the fraction
($N_{\text{confirmed}}$/$N_{\text{selected}}$)
of the selected candidate counterparts based on the selection methods ($N_{\text{selected}}$) that matched the selected candidates confirmed by ALMA ($N_{\text{confirmed}}$).
For the parent sample of 716 $>4\sigma$ robustly detected submillimetre sources in the UDS, we consider only the sources that are covered by both radio and
infrared/optical data (Class 1, as defined in C16).
Applying the radio+OIRTC selection technique on these 523 Class 1 sources, C16 found counterparts for $80\%$ of these (421/523),
of which $37\pm3\%$ have multiple counterparts. Sources for which C16 found no counterparts are not considered in this analysis.
Adding all the Class 1 counterparts (including the multiple counterparts) together,
the selection technique extracted a total of 610 SMG counterparts from 421 submillimetre sources,
which we use for the rest of our analysis. We have repeated our clustering analysis using only one counterpart for each source
(except for those that we have no counterparts for), finding consistent results within the uncertainties. While the clustering does change on small scales,
we determine halo masses by fitting to the large-scale clustering only (see Section 3).

We apply additional quality cuts on the samples. A maximum limit on minimum $\chi^2$ values obtained in
photometric redshift fitting can be used to increase the reliability of source redshifts and to reduce
contamination. A stellar mass completeness cut can also be applied to ensure that a proportion of galaxies with stellar masses above some
limit would be detected. We present clustering measurements of SMGs with photometric redshift $\chi^2<20$ fits, which remove $8\%$ of the SMG sample.
We do, however, note that SMG stellar masses are also uncertain, as the constraints on stellar masses from dusty SEDs is poor. %%%%%%%%%%%%%%%%%%%%%%%%
We apply no mass cut on the SMGs however; most of the SMGs appear to be massive with little dependence on redshift. Overall, $\sim85\%$ of SMGs have stellar masses above
the $90\%$ mass completeness limit (at $z=3.5$) imposed on the passive and star-forming galaxies (M$_{*}>10^{10.2}$M$_{\odot}$, see
Section 2.3). Repeating our clustering analysis after imposing this mass cut on SMGs yielded little difference in our clustering measurements, albeit with slightly larger
uncertainties. Hence we believe that making the mass cut is superfluous and we
choose to keep all $\chi^2<20$ SMGs irrespective of their stellar mass for our analysis.
 We have also repeated our clustering measurements using various cuts in $\chi^2$, $K$-band magnitude and redshifts.
 Adopting more conservative cuts, we found consistent results, while
the size of the errors of our measurements increased due to reduced numbers in each redshift interval.

To test the redshift dependence of our clustering measurements, we split our sample by their redshifts. We present the redshift probability
distribution of our samples in Figure 1. To ensure a reasonable number of SMGs per redshift bin we use intervals of width $\Delta z=0.5$, spanning the redshift
range $1.0<z<3.5$ (see Table 1). We note that the number of SMGs given in Table 1 account for the fact that galaxies
can be represented multiple times across different redshifts as previously described, each with a weight determined by
their redshift probability distributions. The sums of these weights, representing the expected number of galaxies for a given redshift interval, are also listed in Table 1.

\begin{figure}
\centering
\includegraphics[height=0.5\textwidth]{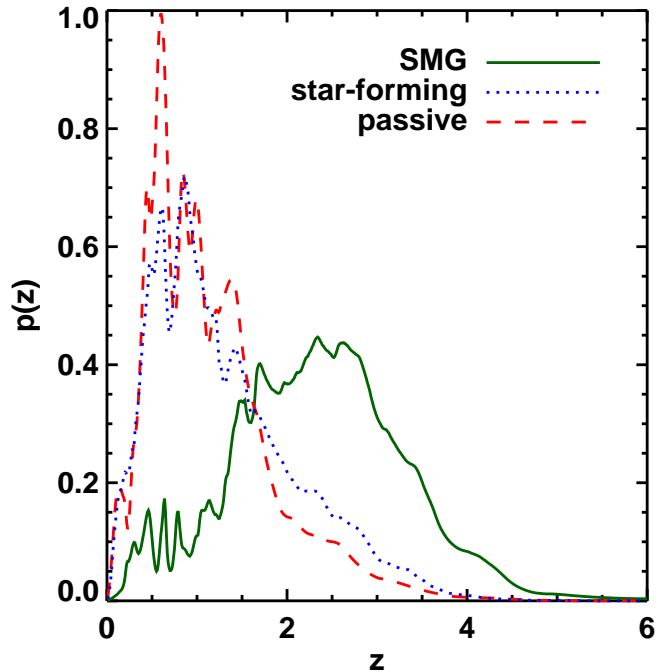}
\vspace{-0.5cm}
\caption{The redshift probability distributions for submillimetre, star-forming and passive galaxies.
 The SMGs are mostly located at redshifts $1.0<z<3.5$, influencing our choice of redshift intervals used in the cross-correlation analysis.
\label{fig:pz}}
\vspace{-0.4cm}
\end{figure}

\vspace{-0.3cm}
%%%%%%%%%%%%%%%%%%%%%%%%%%%%%%%%%%%%%%%%%%%%%%%%%%%%%%%%%%%%%%%%%%%%%%%%%%%%%%%%%%%%%%%%%%%%%%%%%%%%%%%%%%%%%%%%%%%%%%%%%%%%%%%%%%%%%%%%%%%%%%%%%%%%%%%%%%%%%%%%

\subsection{Sample selection - passive and star-forming galaxies}

To enable us to place our clustering results in the context of galaxy evolution, we compare the SMG clustering to
that of the star-forming and passive galaxies at the same redshifts. Taking the parent sample (after separating out the SMGs), we select the comparison populations
using the $U$, $V$ and $J$-Bessel band rest-frame luminosities and subject them to colour criteria that is similar to those derived by \citet{2009ApJ...691.1879W}
and \citet{2013MNRAS.431.3045H}. In order to obtain cleaner samples of star-forming and passive galaxies, we refine our
selection method by choosing quiescent galaxies that were classified by both the $UVJ$ method and an independent PCA method \citep[]{2014MNRAS.440.1880W}.
We shift the $UVJ$ demarcation line from \citet{2013MNRAS.431.3045H} to select passive galaxies as follows:

\begin{align*}
(U - V) &> 0.88 \times (V - J) + 0.89   &       (z&<0.5) \\
(U - V) &> 0.88 \times (V - J) + 0.79   &   (0.5<z&<1.0) \\
(U - V) &> 0.88 \times (V - J) + 0.69   &       (z&>1.0)
\end{align*}

\noindent
with $U - V > 1.5$ and $V - J < 1.4$ at all redshifts. We repeated our clustering analysis using the original selection in \citet{2013MNRAS.431.3045H},
finding similar results, albeit with marginally less separation of passive and star-forming clustering amplitudes. While this criterion correctly identifies most
of these galaxies, the passive sample is still likely to be contaminated by dusty star-forming galaxies that appear to be red and also galaxies hosting AGNs. To
identify dusty contaminants (those that have not already been identified as SMGs),
we use 24\,$\umu$m data and apply the methodology of \citet{2013MNRAS.431.3045H}. Passive galaxies with specific star-formation rate
SSFR$_{24\,\umu m} > 7.43 \times 10^{-11}$\,yr$^{-1}$, composing $3\%$ of $UVJ$-selected passive galaxies, are reassigned to the star-forming sample.

We sort our comparison samples into the same redshift intervals used for the SMG sample and imposed the same quality cuts to further clean our subsamples;
galaxies with $\chi^2 > 20$ (from photometric redshift fits)
were excluded to remove those that cannot be assigned a reliable redshift. Applying a $K$-band magnitude cut of $K = 24.3$ on our passive and star-forming samples, we impose a mass cut
of M$_{*}>10^{10.2}$\,M$_{\odot}$, the $90\%$ mass completeness
limit (up to $z=3.5$). This cut ensures we are comparing passive and star-forming galaxies
of the same stellar mass across all redshift intervals.
We derive this limit using the methodology of \citet{2010A&A...523A..13P} and the implementation described in \citet{2013MNRAS.431.3045H}.
We fit a second order polynomial to the redshift-dependent mass limits (M$_{\text{lim}} =-0.13z^2+1.07z+8.00$, in log solar mass units).
For the volume-limited tracer galaxy population used in the cross-correlation analysis, we apply this redshift-dependent mass completeness limit and reject galaxies 
with stellar masses below M$_{\text{lim}}$.
We caution that the $UVJ$ selection becomes increasingly unreliable for $z > 3$, making passive galaxies difficult to identify correctly. Therefore,
for the highest redshift interval, we combine the passive and star-forming samples together into a combined sample of 1018 galaxies. We use these samples to compare the
clustering of SMGs with those of typical star-forming and passive galaxies within the same redshift intervals. We give details of the various samples in Table 1.

%%%%%%%%%%%%%%%%%%%%%%%%%%%%%%%%%%%%%%%%%%%%%%%%%%%%%%%%%%%%%%%%%%%%%%%%%%%%%%%%%%%%%%%%%%%%%%%%%%%%%%%%%%%%%%%%%%%%%%%%%%%%%%%%%%%%%%%%%%%%%%%%%%%%%%%%%%%%%%%%

\section{CLUSTERING ANALYSIS}
\label{sec:analysis}

To analyse the clustering properties of galaxy populations, we evaluate the 2-point auto-correlation function (ACF). Because we detect 
galaxies on a 2D projected surface, we use the angular version of this function,
a projection of the 3-dimensional spatial correlation function \citep{1980lssu.book.....P}. The ACF
provides us a robust way of tracing the dependence of large-scale structure on galaxy 
properties and evolution through redshift.

The ACF, $w(\theta)$, is a measure of the excess 
probability, compared with a random distribution, of finding a galaxy at an angular 
separation $\theta$ from another galaxy. We use the \citet{1993ApJ...412...64L} estimator, described by

\begin{equation}
w(\theta)=\frac{DD(\theta)-2DR(\theta)+RR(\theta)}{RR(\theta)},
\label{eq:LS}
\end{equation}

\noindent
where $DD(\theta)$, $DR(\theta)$ and $RR(\theta)$ are the galaxy-galaxy, galaxy-random and 
random-random normalised pair counts, respectively. 

We note that our observed field is finite in size, which can lead to an underestimation 
of the clustering by a factor that is defined as the integral constraint. 
We use the formalism of \citet{1999MNRAS.307..703R},

\begin{equation}
C=\frac{\sum RR(\theta).w(\theta)}{\sum RR(\theta)},
\label{eq:IC}
\end{equation}

\noindent
which is dependent on the intrinsic clustering of galaxies, normally by adopting some form for 
$w(\theta)$. Following the method of \citet{2013MNRAS.431.3045H}, we assume $w(\theta)$ to be the 
angular correlation function of the underlying dark 
matter distribution traced by galaxy populations.

To obtain the dark matter correlation function, we determine the non-linear power spectrum, using the formalism 
of \citet{2003MNRAS.341.1311S}. Using the redshift probability distribution, $p(z)$, 
of the galaxy population, we Fourier transform the spectrum and project it to 
calculate the dark matter correlation function.
We then compute the galaxy bias, a measure 
of how well galaxies trace the dark matter distribution. This is quantified by the relationship,

\begin{equation}
w_{\rm{obs}}(\theta)=b^2\times w_{\rm{dm}}(\theta),
\label{eq:wobs}
\end{equation}

\noindent
where $w_{\rm{obs}}$ and $w_{\rm{dm}}$ are the correlation functions of the observed galaxy population 
and dark matter distribution, respectively. The fitting of the dark matter correlation function to 
the observed galaxy correlation function is done by minimising the $\chi^2$. Galaxies 
having a large bias are more likely to be found near the highest density peaks in the dark matter 
mass distribution. 

The galaxy bias is dependent on the dark matter halo mass and the galaxy 
formation epoch, as described by \citet{1996MNRAS.282..347M}. At large scales, $w_{\rm{dm}}$ and $w_{\rm{obs}}$ are consistent 
in both the linear ($w_{\rm{linear}}$) and non-linear ($w_{\rm{non-linear}}$) correlation regimes, where the ACFs are well described by linear gravity theory. At small scales however, they 
deviate; non-linear effects become more significant and we can no longer assume that 
the galaxy population traces the dark matter distribution. We adopt the transition limit between the linear and non-linear scales,
$w_{\rm{non-linear}} < 3\times w_{\rm{linear}}$, in order to constrain the 
fitting of both the integral constraint and the bias at large scales. We also apply an upper limit of $\theta=0.4$ degrees of separation
to this fit, beyond which our measurements become unreliable due to the reduced area resulting from our finite field of view.

Attempting to derive the ACF on small 
data sets is likely to produce large statistical errors, reducing our ability 
to derive well-constrained halo masses. Since we have a small sample of submillimeter galaxies 
at $1.0<z<3.5$, the ACF is not sufficient for a reliable clustering 
analysis on our sample (see C16).

However, we can apply a closely related correlation function: 
the 2-point cross correlation function (CCF), using the positions of the much larger sample of $K$-band selected galaxies. We cross-correlate the desired sample 
population ($D_{\rm{s}}$) with a full volume-limited 90\% mass-complete tracer population ($D_{\rm{t}}$), as follows:

\begin{equation}
w(\theta)=\frac{D_{\rm{s}}D_{\rm{t}}(\theta)-D_{\rm{s}}R(\theta)-D_{\rm{t}}R(\theta)+RR(\theta)}{RR(\theta)}.
\label{eq:cross}
\vspace{0.15cm}
\end{equation}

\noindent
where both data sets are normalised by the total pair counts. By cross-correlating a small target sample (SMGs) with a large tracer population (normal $K$-selected galaxies), 
the cross-correlation function greatly increases the number of pairs, achieving
greatly reduced statistical uncertainties, compared to the auto-correlation. Because each galaxy's contribution to a redshift slice is weighted by its probability distribution
between the given redshift interval limits (see section 2.1), we must apply a weighting scheme to the pair counts. Instead of using a simple pair count, the pair counts
we calculate are the sum of the products of
these weights over pairs of objects. Evaluating the ACF of the large tracer population 
and the CCF of the target sample crossed with the tracer, we inferred the bias of the tracer 
population ($b_{\rm{t}}$) and the ccf bias, $b_{\rm{st}}$ (using equation~\ref{eq:wobs}). Finally, the bias of the sample 
population, $b_{\rm{s}}$, is calculated:

\begin{equation}
b_{\rm{s}}=\frac{b^2_{\rm{st}}}{b_{\rm{t}}}.
\label{eq:bias}
\end{equation}

We infer the ACF of the target sample population 
by multiplying out the tracer population bias from the ccf by 
$(b^2_{\rm{ccf}}/b^2_{\rm{tracer}})$, allowing us to compare the ACF of all sample populations. We ensure that the sample size of the random points is at least 10 times larger than
that of the tracer population.
The errors derived from the ACF and 
CCF are estimated using a bootstrap analysis with 100 repetitions, by resampling the galaxy population with replacement.
We calculate $w(\theta)$ for each of the 100 bootstrap samples and
the derived the variance on $w(\theta)$.
We have made the assumption 
that both the sample and the tracer populations trace the same dark matter distribution 
and that both are linearly biased.

Finally, we estimate the halo masses of our samples using the \citet{2002MNRAS.336..112M} formalism. We also estimate the correlation length, $r_{0}$,
using the formalism of \citet{1980lssu.book.....P}. This quantity is related to the bias parameter as follows:

\begin{equation}
r_{0}=8\bigg(\frac{\Delta_{8}^{2}}{C_{\gamma}}\bigg)^{\frac{1}{\gamma}}
\label{eq:r0}
\end{equation}

\noindent
where $\Delta_{8}$ is the clustering strength of halos more massive than stellar mass $M$ at redshift $z$, defined as $\Delta_{8}=b(M,z)\sigma_{8}D(z)$. The function $D(z)$, as defined by $D(z)=g(z)/[g(0)(1+z)]$, measures the growth factor of linear fluctuations in the dark matter distribution \citep[see, for example,][]{2002MNRAS.336..112M}, with 

\begin{equation}
g(z)=\frac{5}{2}\Omega_{\rm{m}}[\Omega_{\rm{m}}^{4/7}-\Omega_{\rm{\Lambda}}+(1+\Omega_{\rm{m}}/2)(1+\Omega_{\rm{\Lambda}}/70)]^{-1}.
\end{equation}

To maintain a reliable clustering estimate and ensure robust conclusions, we impose a lower limit of sample sizes required for the analysis; if a subsample
for a given redshift interval contains fewer than 30 objects, we do not include it in the analysis presented here.

%%%%%%%%%%%%%%%%%%%%%%%%%%%%%%%%%%%%%%%%%%%%%%%%%%%%%%%%%%%%%%%%%%%%%%%%%%%%%%%%%%%%%%%%%%%%%%%%%%%%%%%%%%%%%%%%%%%%%%%%%%%%%%%%%%%%%%%%%%%%%%%%%%%%%%%%%%%%%%%%

\section{RESULTS AND DISCUSSION}
\label{sec:results}

\subsection{Clustering of $1<z<3$ SMGs and comparison to previous literature}
Previous studies of SMG clustering have focussed on the
broad redshift range $1<z<3$ \citep{2003ApJ...582....6W, 2004ApJ...611..725B, 2009ApJ...707.1201W, 2011ApJ...733...92W, 2012MNRAS.421..284H}. The majority of
SMGs in our sample also lie in this key redshift range, as shown in Figure 1. 
To directly compare with previous clustering measurements,
we first apply our clustering methodology (as described in Section 3)
on the 365 SMGs ($\Sigma_{\text{weight}}=284$) located in the redshift range $1<z<3$.
Cross-correlating all the SMGs in the redshift range $1<z<3$ with a $90\%$ mass-complete tracer sample (normal $K$-selected galaxies)
yielded the galaxy bias value $b=2.18\pm0.97$ and correlation length $r_{0}=4.1^{+2.1}_{-2.0}$h$^{-1}$\,Mpc.
 Using the formalism of \citet{2002MNRAS.336..112M}, our result corresponds to a halo mass of M$_{\text{halo}} \sim 10^{12}$\,M$_{\odot}$.
We plot this in Figure 2, with the accompanying correlation functions, comparing with various previous measurements.

\begin{figure}
\includegraphics[height=0.8\textwidth]{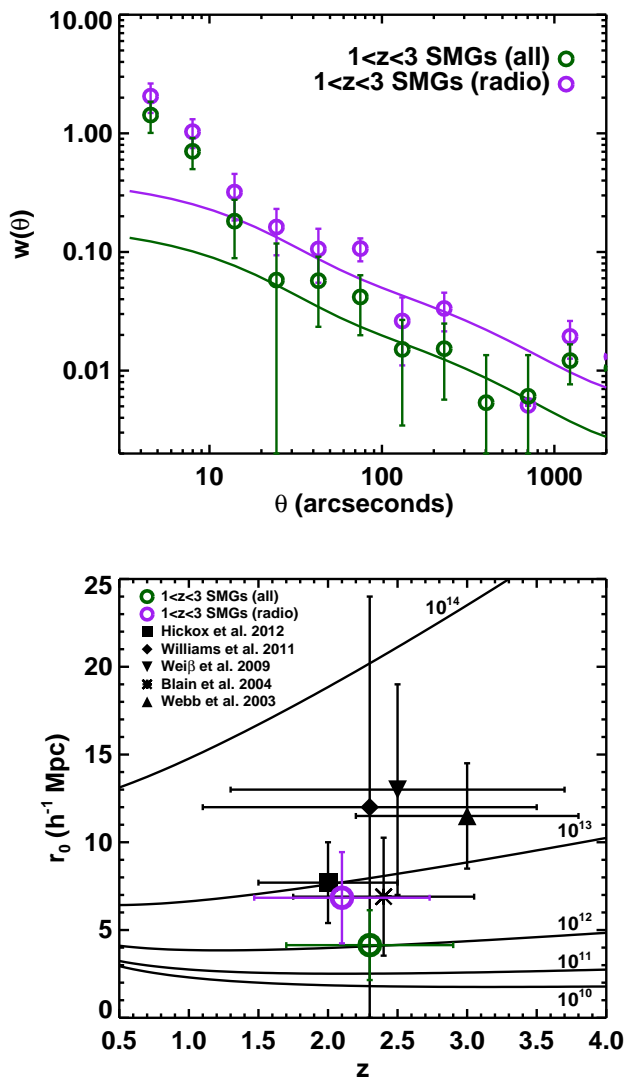}
\caption{\textit{Top panel}: The correlation functions for all $1<z<3$ submillmetre galaxies and the subset of radio-detected submillimetre galaxies. The correlation functions are determined 
by multiplying their corresponding cross correlation functions by $(b^2_{\rm{CCF}}/b^2_{\rm{tracer}})$, where $b$ is the 
galaxy bias. The solid lines are dark matter 
correlation functions fitted onto the observed galaxy correlation functions.
\textit{Bottom panel}: The clustering strength $r_{0}$ as a function of redshift. Open green and purple points are clustering measurements of all the SMGs and SMGs identified
with the radio counterparts, respectively, at redshifts $1<z<3$. The black points are clustering results from previous studies:
\citet{2003ApJ...582....6W, 2004ApJ...611..725B, 2009ApJ...707.1201W, 2011ApJ...733...92W, 2012MNRAS.421..284H}. The curves represent the predicted clustering
strengths for dark matter halos of varying masses (labelled, in solar masses), produced
using the formalism of \citet{2002MNRAS.336..112M}.
\label{fig:prevlit}}
\end{figure}

From Figure 2, we find that our clustering measurement is low, but consistent within the errors, when compared to previous results.
In particular, our result
shows good agreement with that of \citet{2004ApJ...611..725B} (corrected by \citealt{2005ApJ...619..697A}) and \citet{2012MNRAS.421..284H}.
The consistency with \citet{2004ApJ...611..725B} improves further if we compare with their reported alternative
clustering strength ($r_0=5.5\pm1.8$h$^{-1}$\,Mpc) in the absence of the SSA22 field, the most overdense field surveyed in their investigation and a well-known proto-cluster region
\citep{1998ApJ...492..428S}.
We note that our sample is $\sim5\times$ larger than that of \citet{2012MNRAS.421..284H}, which used the largest SMG sample in a similar analysis prior to this study.
Our measured clustering strengths are also consistent with previous clustering studies of infra-red galaxies selected at shorter wavelengths
\citep[e.g.,][where the latter three studied SMGs in the Herschel field]{2006ApJ...641L..17F,Cooray+2010,2010A&A...518L..11M,2012ApJ...753...23M}.

It appears at first sight that, on average, SMGs are not as strongly clustered as previously thought.
 Alternatively, the SMG clustering could be more complex and may be dependent on redshift, large-scale environment and merger history.
\citet{2009ApJ...691..560C} discussed the dependence of a complex bias on the large-scale environment and merger history.
 They proposed that SMGs may reside in smaller halos than would be inferred from the linear bias model
assumed in the halo modelling for this study. In that case, SMGs do not necessarily trace the most massive dark matter halos in the Universe.

The relatively low clustering measurement we derive may be affected by the complex nature of SMG clustering that evolves with redshift. Limiting our redshift interval
size to $1.5<z<3.0$ and cross correlating the resulting sample of 327 SMGs ($\Sigma_{\text{weight}}=244$), we find SMGs are more strongly clustered, with a bias $b=4.40\pm0.86$,
 in better agreement with previous studies. This stronger clustering signal
suggests that the excluded low redshift SMGs are weakly clustered and they are diluting the stronger clustering exhibited by higher
redshift SMGs. This indicates a possible redshift evolution in SMG clustering, which we explore in detail in Section 4.2.

Previous studies had small sample sizes, typically comprising the more luminous (radio-identified) SMGs.
In contrast, our sample of SMGs includes many fainter counterparts and a significant fraction ($\sim50\%$) having no strong radio emission. Therefore it is possible that
previous clustering measurements were biased towards the brightest radio SMGs, which may be a more extreme luminous subset of the SMG population.
To test this claim, in Figure 2, we present the clustering measurement of SMGs that have radio counterparts only. We find
radio SMGs to be slightly more strongly clustered with respect to all SMGs, with bias $b = 3.20\pm1.12$, correlation length $r_{0}=6.8^{+2.7}_{-2.6}$h$^{-1}$\,Mpc and halo mass
M$_{\text{halo}}\sim 10^{13}$\,M$_{\odot}$. This measurement is also consistent with previous studies, in good agreement with \citet{2004ApJ...611..725B} and
\citet{2012MNRAS.421..284H}, within $1\sigma$ errors.
The improved consistency suggests that the higher luminosity radio SMGs previously studied tend to 
reside in higher mass halos, as was the consensus in the previous studies. Investigating the possibility of the clustering dependence
 on radio emission and S$_{850\,{\umu}\text{m}}$ flux density further would require a much larger sample of SMGs than even we have presently.
We attempted to measure the clustering strength
as a function of S$_{850\,{\umu}\text{m}}$ flux density for the SMG sample at redshifts $1<z<3$. We found no significant
trend, however, since the uncertainties were too large.

Finally, it is worth noting that clustering measurements performed with single-dish surveys are
subjected to a blending bias \citep[e.g.][]{2013ApJ...768...91H,2013MNRAS.432....2K,2015arXiv150404516C}.
This describes the contribution to the clustering signal due to the blending of SMGs into single submillimetre sources as a result of the low resolution. We discuss this in detail in section
4.3. Summarising briefly, \citet{2015arXiv150404516C} simulated this effect at $850\,\umu$m to match observations in the SCUBA-2 map. They suggest
that the confusion between the sources can artificially increase the galaxy bias measurements by a factor of $\sim 2$ and that
any galaxy bias measured with a single dish survey must be corrected for by this factor. We note that the factor $\sim 4$ derived in \citet{2015arXiv150404516C} applies to the
clustering amplitude, $A$, for which this quantity scales with the galaxy bias squared $b^2$. In addition, surveys with larger beams are subjected to a greater blending
bias.
The \citet{2012MNRAS.421..284H} and \citet{2009ApJ...707.1201W} clustering measurements were carried out with the ECDFS LABOCA survey,
which has a FWHM beam of $\sim20''$. Similarly \citet{2011ApJ...733...92W} utilised the 28$''$ AzTEC/ASTE beam,
selecting SMGs at $1100\,\umu$m.
The clustering measurements in the previous literature could therefore be subjected to a larger correction than the value derived for the current SCUBA-2 survey.
It is likely that correcting for blending would bring previous studies into better agreement with the clustering measurements presented here.

%%%%%%%%%%%%%%%%%%%%%%%%%%%%%%%%%%%%%%%%%%%%%%%%%%%%%%%%%%%%%%%%%%%%%%%%%%%%%%%%%%%%%%%%%%%%%%%%%%%%%%%%%%%%%%%%%%%%%%%%%%%%%%%%%%%%%%%%%%%%%%%%%%%%%%%%%%%%%%%%

\subsection{Evolution of SMG halo masses with redshift}
Having a large sample of SMGs in a single field, we can now test the redshift dependence of SMG clustering derived using a cross-correlation technique.
We also compare the clustering measurements with those of the $UVJ$-selected passive and star-forming galaxies, to reveal their possible relation to SMGs.
Splitting the sample into the redshift intervals described in Section 2.2 and applying the cross-correlation analysis in Section 3, we plot
the resulting correlation functions in Figure 3. We compare the 
clustering of SMGs, passive and star-forming galaxies in each of the redshift intervals.
As noted earlier, in the highest redshift interval ($3.0<z<3.5$), we combine the passive and star-forming objects into a single sample containing both
populations as we can no longer rely on the $UVJ$ selection at such high redshifts.
For clarity, we plot error bars for the SMG correlation functions only. The auto-correlation functions are derived from their respective 
cross-correlation functions multiplied by $(b^2_{\rm{CCF}}/b^2_{\rm{tracer}})$.

\begin{figure}
\centering
\vspace{-0.3cm}
\includegraphics[height=1.05\textwidth]{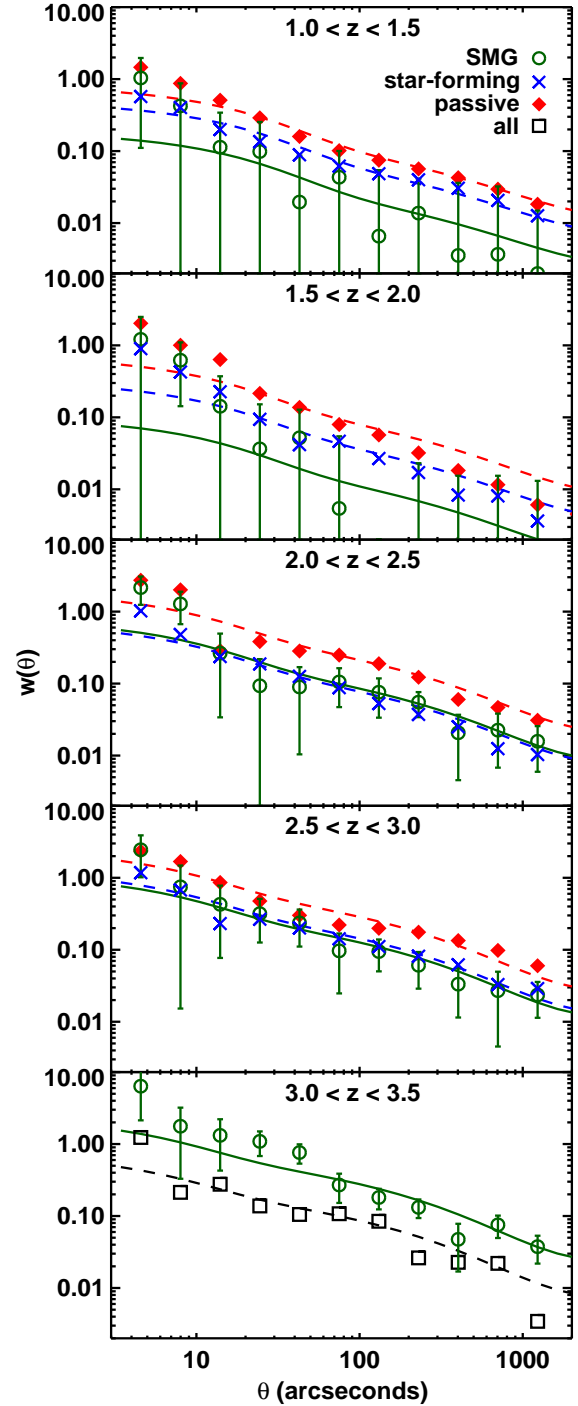}
\caption{The correlation functions for submillmetre galaxies, compared with comparison samples of passive
and star-forming $K$--band selected galaxies. The correlation functions are determined 
by multiplying their corresponding cross correlation functions by $(b^2_{\rm{CCF}}/b^2_{\rm{tracer}})$, where $b$ is the 
galaxy bias. The panels represent redshift intervals as marked. The solid (dashed) lines are dark matter 
correlation functions fitted onto the observed SMG (passive and star-forming) correlation functions. For $3.0<z<3.5$,
the black squares represent all detected galaxies minus SMGs.
The error bars (typically of order $\sim0.01$) for the non-SMG samples have been omitted for clarity.
\label{fig:corr}}
\end{figure}

From the 2D (projected on sky) representation of clustering, we find that passive galaxies exhibit stronger clustering amplitudes with respect to their
star-forming counterparts, at all redshifts up to $z=3$. This is in good agreement with previous studies, including \citet{2013MNRAS.431.3045H},
who also studied the UDS field. SMGs, however, appear to have a range of clustering amplitudes in different
redshift intervals. At $1<z<2$, SMGs appear to be weakly clustered, consistent with their normal star-forming counterparts. At $z>2$, the SMG correlation functions
are consistent with those of both the passive and star-forming samples, within $1\sigma$. The results for the redshift range $3.0<z<3.5$ suggest
that the highest-redshift SMGs are tentatively more clustered than normal galaxies, on average.

\begin{table*}
\centering
\begin{tabular}{ |c c c c c c c c c| }
\hline
 N$_{\text{gal}}$ & $\Sigma_{\text{weight}}$ & $z_{\text{min}}$ & $z_{\text{max}}$ & $b$ & $\sigma_{b}$ & $r_0$ & bb & $b_{\text{corr}}$ \\
\hline
\textbf{sub-millimetre} \\
61 & 39.95 & 1.0 & 1.5 & 1.20 & 0.89 & $3.19^{+2.72}_{-2.48}$ & 1.28 & 0.94 \\
127 & 76.50 & 1.5 & 2.0 & 1.03 & 0.96 & $2.19^{+2.36}_{-2.08}$ & 1.18 & 0.87 \\
172 & 82.74 & 2.0 & 2.5 & 3.86 & 1.07 & $7.98^{+2.48}_{-2.41}$ & 1.15 & 3.36 \\
176 & 84.31 & 2.5 & 3.0 & 5.00 & 1.21 & $9.08^{+2.47}_{-2.41}$ & 1.23 & 4.07 \\
127 & 50.77 & 3.0 & 3.5 & 8.81 & 2.75 & $14.87^{+5.24}_{-5.06}$ & 1.22 & 7.22 \\
\hline
\textbf{radio sub-millimetre} \\
51 & 36.57 & 1.0 & 1.5 & 1.10 & 1.01 & $2.89^{+3.07}_{-2.71}$ & 1.28 & 0.87 \\
70 & 45.72 & 1.5 & 2.0 & 1.65 & 1.09 & $3.69^{+2.79}_{-2.58}$ & 1.18 & 1.40 \\
78 & 43.55 & 2.0 & 2.5 & 2.81 & 1.48 & $5.59^{+3.15}_{-3.15}$ & 1.15 & 2.44 \\
64 & 31.46 & 2.5 & 3.0 & 5.93 & 2.14 & $10.98^{+4.48}_{-4.30}$ & 1.22 & 4.86 \\
\hline
\textbf{star-forming} \\
706 & 540.61 & 0.0 & 0.5 & 0.53 & 0.15 & $2.16^{+0.69}_{-0.67}$ & - & - \\
2208 & 1774.09 & 0.5 & 1.0 & 0.88 & 0.13 & $2.87^{+0.48}_{-0.47}$ & - & - \\
2664 & 1854.41 & 1.0 & 1.5 & 1.92 & 0.14 & $5.37^{+0.44}_{-0.43}$ & - & - \\
2231 & 1351.12 & 1.5 & 2.0 & 1.81 & 0.21 & $4.09^{+0.53}_{-0.52}$ & - & - \\
1840 & 914.25 & 2.0 & 2.5 & 3.69 & 0.37 & $7.56^{+0.85}_{-0.84}$ & - & - \\
1275 & 639.00 & 2.5 & 3.0 & 5.25 & 0.52 & $9.59^{+1.06}_{-1.05}$ & - & - \\
\hline
\textbf{passive} \\
480 & 398.07 & 0.0 & 0.5 & 0.95 & 0.22 & $4.13^{+1.07}_{-1.05}$ & - & - \\
2160 & 1809.48 & 0.5 & 1.0 & 1.83 & 0.14 & $6.48^{+0.55}_{-0.55}$ & - & - \\
2219 & 1678.81 & 1.0 & 1.5 & 2.45 & 0.15 & $7.04^{+0.48}_{-0.48}$ & - & - \\
1423 & 936.83 & 1.5 & 2.0 & 2.60 & 0.25 & $6.12^{+0.66}_{-0.65}$ & - & - \\
739 & 350.28 & 2.0 & 2.5 & 6.24 & 0.59 & $13.55^{+1.43}_{-1.42}$ & - & - \\
531 & 228.49 & 2.5 & 3.0 & 7.69 & 0.74 & $14.65^{+1.57}_{-1.56}$ & - & - \\
\hline
\textbf{star-forming+passive} \\
1018 & 457.22 & 3.0 & 3.5 & 5.01 & 1.09 & $7.94^{+1.94}_{-1.89}$ & - & - \\
\hline
\end{tabular}
\caption{Table of clustering measurements used in Figure 4. The columns shown are: number of galaxies, the sum of weights (the expected number of galaxies from the redshift probability
distributions), redshift intervals, galaxy bias (plotted in Figure 4),
1$\sigma$ uncertainty on the bias, the clustering strengths (in units of h$^{-1}$\,Mpc) and their 1$\sigma$ uncertainties,
estimated blending bias (bb) for SMGs (see section 4.3), SMG galaxy bias corrected for blending bias.}
\label{table:1}
\end{table*}

\begin{figure*}
\centering
\includegraphics[height=0.6\textwidth]{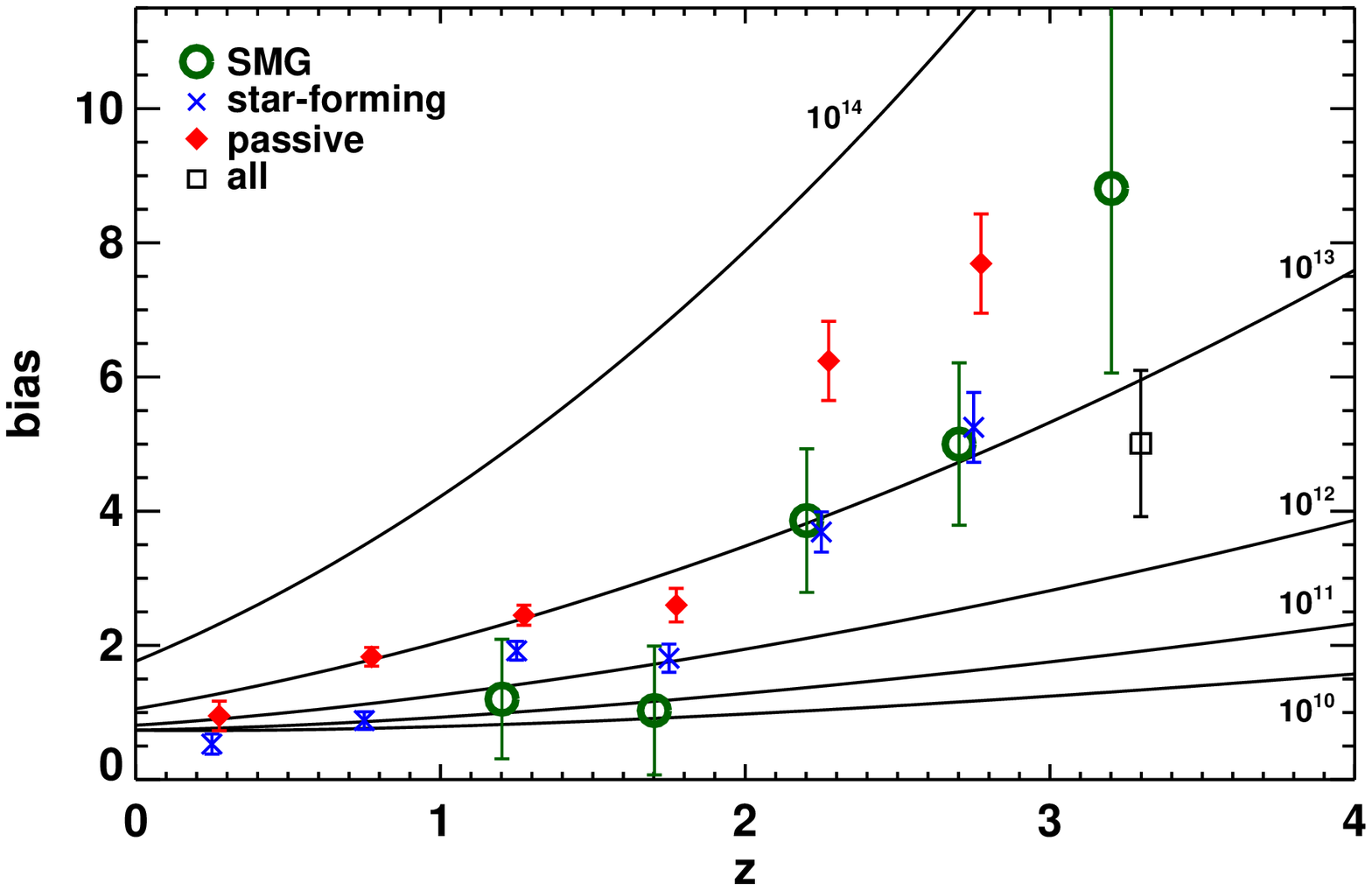}
\caption{Galaxy bias versus redshift for submillimetre galaxies (open green circles), compared with a sample of $K$-band passive and star-forming galaxies
(filled red diamonds and blue crosses, respectively). The points are offset 
slightly for clarity. The solid lines show the evolution of bias for
 dark matter halos (produced using the formalism of \citealt{2002MNRAS.336..112M}), with varying mass (labelled, in solar masses).
\label{fig:halomass}}
\end{figure*}

For a more intuitive representation of the clustering results, in Figure 4 we plot the bias measurements computed from the cross-correlation methodology in
Section 3 and relate them to halo masses using the \citet{2002MNRAS.336..112M} formalism. We also list the bias measurements and their uncertainties in Table 1.
From Figure 4, a number of trends are apparent. First, we find that passive galaxies reside in dark matter halos of 
mass $\sim 10^{13}$\,M$_{\odot}$, consistent across most redshift intervals (in good agreement with \citealt{2013MNRAS.431.3045H}).
On the other hand, star-forming galaxies are clearly more weakly clustered than their
passive counterparts, residing in low mass halos ($\sim 10^{10}$--$10^{12}$\,M$_{\odot}$) up to redshift $z\sim2$.
At higher redshifts, the star-forming galaxies appear to reside in halos of increasingly higher mass. Thus star-forming galaxies show evidence for downsizing,
with activity shifting towards lower mass halos as the Universe ages.
Our expectations of 
the relative clustering between passive and star-forming objects are confirmed, with our results reflecting the findings 
of numerous previous studies \citep[e.g.][]{2006A&A...452..387M, 2009ApJ...691.1879W, 2011ApJ...727..111F, 2011ApJ...728...46W, 2012ApJ...750...37J, 2013MNRAS.431.3045H}.

To place these clustering results in context, we discuss a common feature in many galaxy evolution models, the hot halo scenario, which suggests 
that galaxy evolution is dependent on dark matter halo mass
\citep[e.g.][]{2006MNRAS.367..864C, 2011ApJ...741...99C, 2015MNRAS.449..901M, 2016ApJ...817...97L}. 
A dark matter halo growing in mass becomes more effective in maintaining a reservoir of gas 
with cooling times longer than the age of the Universe (e.g. a few Gyrs). Any additional gas infalling into the halo
is shock heated and is prevented from forming stars. The virial temperature 
of the dark matter halo will continue to increase from the accretion of more halos while reservoirs of cold gas within any galaxy in the halo become exhausted 
and star formation rates in galaxies decline as they can no longer accrete. Galaxies that reside in more massive halos will have their 
star formation increasingly quenched. This suggests that there is some threshold halo mass 
limit above which galaxies become progressively passive. According to recent models, this 
limit is of order few $\sim10^{12}$M$_{\odot}$ with maximum quenching at $10^{13}$M$_{\odot}$ 
\citep{2006MNRAS.367..864C, 2011ApJ...741...99C}, consistent with the derived halo masses for passive galaxies in this study.

Our population of interest, the SMGs, appear to exhibit a clustering signal that is dependent on redshift; the downsizing effect appears to be even stronger
than seen in star-forming galaxies. The downsizing effect appears to confirm predictions made with the \citet{2013MNRAS.434.2572H} mock SMG catalogues used by \citet{2015MNRAS.452..878M}.
At $1<z<2$, SMGs reside in halos of relatively low masses, $\sim 10^{11}$\,M$_{\odot}$.
This is consistent with star-forming galaxies, thus SMGs are weakly clustered at this epoch with respect to passive galaxies. 
As we advance to higher redshifts ($z>2$), we see a stronger SMG clustering amplitude, although still consistent with star-forming galaxies. We compute
 halo masses M$_{\text{halo}} \sim5.89\times10^{12}$\,M$_{\odot}$ and M$_{\text{halo}}\sim1.26\times10^{13}$\,M$_{\odot}$
 for redshift intervals
$2.0<z<2.5$ and $2.5<z<3.0$, respectively. The results for these redshifts are in reasonable agreement with previous studies
\citep{2003ApJ...582....6W, 2004ApJ...611..725B, 2009ApJ...707.1201W, 2011ApJ...733...92W, 2012MNRAS.421..284H},
and in contrast to the low SMG clustering amplitude at lower redshifts.

Comparing to galaxy populations selected at shorter wavelengths than the 850\,$\umu$m sample used in this study, we find our estimated clustering measurements
 broadly consistent with a number of studies \citep[e.g.,][]{2006ApJ...641L..17F,Cooray+2010,2010A&A...518L..11M,2012ApJ...753...23M}. \citet{2006ApJ...641L..17F}
studied the spatial clustering of galaxies selected in the IRAC bands using a 1.6\,$\umu$m emission feature, with star-formation rates similar to SMGs.
Splitting the sample into two redshift bins between 1.5 and 3.0,
the authors derived halo masses of M$_{\text{halo}} \sim6\times10^{13}$\,M$_{\odot}$, consistent with our measurements at the high redshift bins. However, they reported no
strong redshift evolution in the clustering of their selected galaxy samples.
In contrast, \citet{2013MNRAS.433..127M} found the same halo downsizing trend as reported in this study, having performed a clustering analysis
on galaxies selected at wavelength 60\,$\umu$m (SFR\,$\geq100$\,M$_{\odot}$yr$^{-1}$) in the COSMOS and EGS fields.

The relatively weak SMG clustering seen at redshifts $1<z<2$ demonstrates that the SMGs at this epoch are unlikely to be the
progenitors of the massive ($\sim2$--$4$\,L$^*$) elliptical galaxies we see
in the local Universe. The typical bias measurements, halo masses and hence the environment of these elliptical galaxies
 do not match the measurements of the SMGs at redshifts $1<z<2$. This finding is emphasised in Figure 5, where we plot the expected evolution of the dark matter 
bias for halos with observed M$_{\text{halo}}$ for SMGs. This evolution is calculated using the \citet{2010MNRAS.406.2267F}
formalism of the median growth rate of halos, as a function of halo mass M$_{\text{halo}}$ and redshift.

\begin{figure}
\centering
\includegraphics[height=0.5\textwidth]{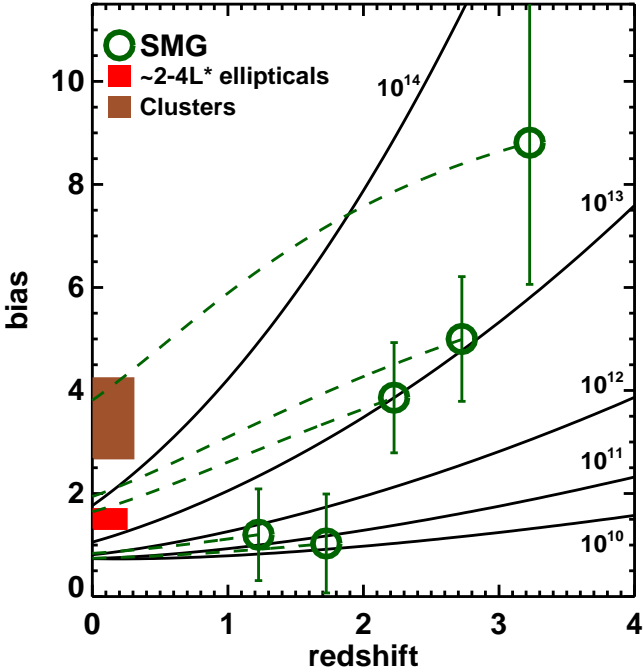}
\caption{Predicted galaxy bias evolution of SMGs (open green circles) at $z>1$,
plotted against redshift. The solid black lines represent the expected bias evolution for dark matter halos with constant masses (labelled, in solar masses).
We trace the mass growth of halos hosting SMGs in green dashed lines, using
the formalism from Fakhouri et al. (2010). We also plot typical bias measurements of $\sim2-4$\,L$^*$ galaxies (from the luminosity-bias relation derived in \citealt{2011ApJ...736...59Z})
and optically selected galaxy clusters at redshifts $0.1<z<0.3$ \citep{2009ApJ...692..265E}.
\label{fig:halomass}}
\end{figure}

Tracing the growth of halos over redshifts $1<z<2$, it is clear that halos hosting SMGs at these redshifts do not evolve to become halos hosting massive passive
galaxies at the present day. This further supports the idea that these SMGs are not the progenitors of the massive elliptical galaxies we see today,
which typically reside in halos with a minimum mass of $\sim 10^{13}$\,M$_{\odot}$ \citep[e.g.,][]{2003MNRAS.346..186M,2011ApJ...736...59Z}.
 However, halos hosting SMGs in the redshift range $2.0<z<2.5$ are
consistent with their low-redshift passive counterparts, emphasising that in order for the SMGs to be these progenitors,
 they must typically form at redshifts $z>2$. It is worth noting that in the highest redshift intervals, $z>2.5$, SMG host halos would eventually evolve into very massive
halos (M$_{\text{halo}}>10^{14}$\,M$_{\odot}$), which typically host galaxy clusters at the present day \citep{2009ApJ...692..265E}. 

We place these findings further into their observational context by commenting on the following two previous studies of SMGs in single clusters at different redshift epochs.
 First, \citet{2014ApJ...782...19S} studied a $z=1.62$ cluster in the UKIDSS SCUBA-2 CLS, hosting a mix of massive passive galaxies and highly
active galaxies with significant dust content. In summary, they found that the most active cluster members reside in the underdense regions in the cluster, while
the dense cores are populated in this structure at this epoch by massive passive galaxies.
Hence massive galaxies in the dense cores of the local Universe were already likely to be in place at $z=1.6$,
and any ultraluminous star-forming galaxies found in this cluster of interest are likely to be infalling onto this structure. Therefore, the progenitors of
these elliptical galaxies are likely to have undergone a transformation at $z>2$.

The halo downsizing phenomenon is further supported when we consider a second study, by \citet{2015ApJ...815L...8U}, of a protocluster traced by Lyman alpha emitters (LAEs) at $z=3.1$.
Using the Astronomical Thermal Emission Camera (AzTEC),
they detected ten SMGs within this protocluster located in the SSA22 field. A cross-correlation technique with LAEs revealed
that SMGs are strongly clustered and reside at the core of this protocluster. Additionally, \citet{2014A&A...570A..55D} found an excess of $870\,\umu$m-selected SMGs
in a protocluster associated with the radio galaxy MRC1138-262 at $z=2.16$.
 If other protoclusters at this epoch host SMGs in their cores, then this
finding would be consistent with the strong clustering amplitudes we measured for $z>2$. Combining results from both \citet{2014ApJ...782...19S} and \citet{2015ApJ...815L...8U}, 
it appears that SMG activity shifts from high density environments at high redshifts to lower density environments at later epochs, as seen in our clustering results.

When considering the link of SMGs to the progenitors of local elliptical galaxies, we must make the following caveat;
we also find normal star-forming galaxies in the same mass halos as the SMGs, at every redshift interval.
Because star-forming galaxies exhibit clustering signals indistinguishable from SMGs, we cannot
rule out the possibility that these normal galaxies at $z>2$ could also evolve to become the local massive elliptical galaxies. Adopting this view, SMGs are just the more
dusty highly star-forming subset of normal star-forming galaxies, found at the top of the main sequence. 
SFRs derived using far infra-red photometry (Micha\l{}owski et al. in prep.) reveal that most SMGs do indeed lie on the main sequence of star-forming
galaxies, with an average SFR evolving with time, from
SFR$_{\text{av}}\sim360$\,M$_{\odot}$yr$^{-1}$ at redshifts $3.0<z<3.5$, to SFR$_{\text{av}}\sim190$\,M$_{\odot}$yr$^{-1}$ at redshifts $1.0<z<1.5$.
In contrast, measuring SFRs from optical/IR photometry of massive star-forming galaxies, we find a mean SFR of $\sim100$\,M$_{\odot}$yr$^{-1}$, which
has no significant evolution with redshift. It is therefore possible that optical/IR determinations are missing the most dust-enshrouded star formation,
but we defer a detailed investigation of the relationship between SFR and halo mass to future work. For the present work, we simply note the striking
similarity between the clustering of SMGs and normal star-forming galaxies, suggesting no strong link between SFR and environment when comparing star-forming galaxies at a given epoch.

An alternative way to compare high-redshift star-forming galaxies with their possible descendants is to consider space densities. The number
density of normal star-forming galaxies is found to be at least $\sim2\times10^{-4}$\,Mpc$^{-3}$, which rises with decreasing redshift (up to $z=1$). Considering that this is larger
than the typical number density of local luminous red galaxies \citep[$\simeq10^{-4}$\,Mpc$^{-3}$, e.g.,][]{2006MNRAS.372..537W}, it is likely that most star-forming galaxies do not
evolve to become these massive red systems by the present day. SMGs have low number densities of $\simeq1-2\times10^{-5}$\,Mpc$^{-3}$, which remain constant across all redshift
intervals considered in this study. By calculating the SMG number densities, we can estimate the number densities of the SMG descendants.
Assuming that each SMG galaxy goes through one SMG phase in its lifetime, we determine the descendant number density, $\phi$, using the simple relation,

\begin{equation}
\phi=\rho_{\rm{SMG}}\times\frac{\tau_{\rm{obs}}}{\tau_{\rm{burst}}},
\label{eq:desc_n}
\end{equation}

\noindent
where $\rho_{\text{SMG}}$ is the observed SMG number density, $\tau_{\text{obs}}$ is the time between each redshift interval considered and $\tau_{\text{burst}}$ is the SMG lifetime.
We note that most of the uncertainty lies with the estimation of the SMG lifetime. Numerous previous studies estimated
the lifetime to be $\sim100$\,Myr, by using gas mass measurements from CO observations in SMGs \citep{2005ApJ...622..772C},
estimating gas consumption time-scales and fading times \citep{2006MNRAS.371..465S},
or employing simple stellar evolution models based on star-formation timescale \citep{2011ApJ...740...96H}.
Adopting $\tau_{\text{burst}}=100$\,Myr implies that the expected number densities will be 
$\phi\simeq4\times10^{-5}$--$2\times10^{-4}$\,Mpc$^{-3}$, for SMGs at redshifts $1.0<z<3.5$.
Interestingly, SMG descendants located at redshifts $2.0<z<2.5$ and $2.5<z<3.0$ are expected to have number densities of $\phi\simeq1\times10^{-4}$\,Mpc$^{-3}$,
in excellent agreement with the number density of luminous red galaxies. These arguments provide further evidence that SMGs located at redshifts $z\simeq2$ are likely progenitors of these
massive elliptical galaxies, while possibly eliminating the normal star-forming galaxies as candidates, on account of their higher number densities.
%%%%%%%%%%%%%%%%%%%%%%%%%%%%%%%%%%%%%%%%%%%%%%%%%%%%%%%%%%%%%%%%%%%%%%%%%%%%%%%%%%%%%%%%%%%%%%%%%%%%%%%%%%%%%%%%%%%%%%%%%%%%%%%%%%%%%%%%%%%%%%%%%%%%%%%%%%%%%%%%

\subsection{Uncertainties}

We discuss here a number of uncertainties in our analysis.
 The construction of our sample could be performed in 
a variety of ways and although we believe we have used the optimal selection method, there is no single correct method.
 The choice of cuts we make with such a small sample of
SMGs could significantly impact our clustering measurements. We have applied a variety of cuts on the $\chi^2$ of the
 photometric redshift fitting and mass, as well as a selection of
different redshift binning methods.
 It is, in essence, a compromise between securing large numbers of SMGs and
maintaining high quality photometric data. Varying the selection makes little qualitative difference to the
conclusions of this work -- in most cases, we see evidence of a downsizing effect on halos hosting SMGs.
 However, our measurement in the lowest SMG redshift interval
($1.0<z<1.5$) is sensitive to the choice of selection. This variation is most likely due to the low number of SMGs in this bin ($\sim60$).
While the clustering amplitudes in the higher redshift bins are relatively stable,
in this bin we find variations in the bias measurement by up to a factor of two, although formally always
consistent with the star-forming galaxies at this epoch (Figure 4). We conclude that larger samples are required for more robust measurements of clustering
in this redshift range.

The low clustering measurements for redshifts $1<z<2$ may be attributed to the reliability of the SMG identifications. 
%We also note that the negative K-correction is weaker at this epoch, as the luminosity measured in the $850\,\umu$m waveband is a factor $\sim2$ lower
%than the luminosity a source would have if it resided at redshifts $z>2$.
Because a lower clustering measurement
corresponds to a increasingly random distribution of objects, this could reflect an increasing contamination in our low-redshift SMG sample.
To test this scenario, we performed a clustering analysis on the SMGs with secure radio counterparts, whose selection technique is well established
in previous literature \citep[e.g.][]{2002MNRAS.337....1I,2005MNRAS.364.1025I,2006MNRAS.370.1185P,2012MNRAS.420..957Y}.
The resulting bias and halo mass measurements, which we present in Table 1, are broadly consistent with those of the radio$+$OIR SMGs,
especially at low redshifts ($1<z<2$). This is no surprise, however, as nearly all SMGs at lower redshifts are identified as radio counterparts (see Figure 1), reflecting the
fact that the radio selection technique is best at picking out the luminous SMGs at low redshifts.
This complements the OIRTC selection, which is better at identifying less luminous SMGs at redshifts
$z>2$. Generally, SMGs selected using radio counterparts are skewed toward lower redshifts; the fraction of SMGs with radio counterparts
 reduces from 84\% at redshifts $1.0<z<1.5$ to 36\% at redshifts $2.5<z<3.0$. %Is this because radio fluxes drop rapidly with redshift, on account of their positive k-correction - therefore harder to assign radio counterparts to high redshift sources?
In fact, based on the ALMA training sample, the OIRTC and radio selection techniques both have the same completeness and accuracy, $52\pm12\%$ and $87^{+13}_{-23}\%$
respectively \citep{2016arXiv160102630C}.
We also note that the stellar mass distribution of radio SMGs is consistent with that of SMGs with combined OIRTC/radio counterparts (i.e. most of these galaxies
have stellar masses M$_{*}>10^{10.2}$\,M$_{\odot}$), with no
significant evolution with redshift. Hence, we are confident that the low-redshift SMGs are not heavily contaminated by misidentified galaxies
and that the low clustering amplitudes appear to reflect the
scenario in which low-redshift SMGs live in lower density environments compared to their high-redshift counterparts.

We also note that our SMG identification is incomplete, having associated counterparts to $80\%$ of the submillimetre sources. Hence there are submillimetre sources whose clustering
properties are not being measured. We remark that most of these sources tend to be fainter, in which the incompleteness fraction increases to $\sim30\%$ for sources with flux
density S$_{850\,{\umu}\text{m}}<$ 5\,mJy. These may be faint sources blending together to produce brighter submillimetre signals above the flux limit of 2.7\,mJy,
but for which no counterparts can be
detected. Alternatively they may be sources located at high redshifts ($z>3$), for which the radio$+$OIR selection techniques are limited in their ability to identify SMGs.
We attempt to measure the effect of the counterpart incompleteness on the clustering signal by calculating the ACF of all $4\,\sigma$ sources and compare to the
ACF of sources for which we have identified counterparts. Both measurements yielded a clustering amplitude $A\sim5$, consistent within 1\,$\sigma$. Hence with regards to the angular
clustering, the incompleteness of the SMG identification has little effect. However, it is hard to determine whether it can influence the bias measurements at different redshifts.

An interesting galaxy population to investigate further are those selected by the OIRTC criteria alone, without bright SMG counterparts.
There are many more ($\sim 3000$)
submillimeter sources, fainter than 2.7\,mJy at wavelength $850\,\umu$m (the detection limit of the S2CLS map), that satisfy the OIRTC criteria. 
A clustering analysis of this population (Chen et al. in prep.) %, though majority of which have fainter 850 flux than our sample,
may reduce the size of the error bars of the clustering amplitudes and provide additional constraints on
predicted halo masses.

We do not take into account the large-scale cosmic variance in our estimation of our errors, other than the internal uncertainty quantified in our
bootstrap analysis. Ideally we would have
 multiple independent fields of SMG identifications, which will be soon possible with forthcoming mapping of the COSMOS field with SCUBA-2 in the East Asian Observatory JCMT S2COSMOS survey.

Finally, we return to the issue that submillimeter sources suffer from a blending bias,
arising from the blending of multiple counterparts within the SCUBA-2 beam \citep[e.g.,][]{2013ApJ...768...91H,2013MNRAS.432....2K}.
\citet{2015arXiv150404516C} investigated the clustering evolution of SMGs selected using a combination of the Millennium-style N-body
simulation, an updated version of \textsc{galform} semi-analytic model of galaxy formation \citep{2015arXiv150908473L} and a model for calculating the absorption and re-emission of stellar
radiation by dust. They found that galaxies selected by their $850\,\umu$m flux densities are hosted by dark matter halos of mass $\sim10^{11.5}$--$10^{12}$h$^{-1}$\,M$_{\odot}$
over a large redshift range of $0<z<4$, consistent with
 our clustering measurements at $1<z<2$. By simulating the sub-mm imaging at $850\,\umu$m to match observations with SCUBA-2,
(using the method outlined in \citealt{2015MNRAS.446.1784C}),
 they demonstrated that confusion between low-resolution sources can artificially boost galaxy bias
measurements by a factor of $\sim 2$.

To investigate whether this will affect the results presented in this work we have repeated the analysis of \citet{2015arXiv150404516C},
using the same redshift bins used in our analysis.
In order to assign a redshift to a submillimetre source identified in the simulated imaging, we use the redshift of the galaxy that contributes the
largest fraction of the detected single-dish sub-millimetre flux to the source.
Calculating the clustering of the sources in the simulated image within the redshift intervals used here and comparing to the
clustering of simulated galaxies in the same redshift interval, we find the blending bias correction reduces from a factor of $\sim2$ to $\sim1.2$.
We attribute this to a reduction in the beam-induced correlations between SMGs at different redshifts as discussed in \citet{2015arXiv150404516C}. 
We list these corrections and the corrected galaxy bias
values in Table 1, but we have not applied these corrections in the figures. The corrected SMG clustering measurements are lower than (or consistent with, within 1$\sigma$ errors)
 those of the star-forming sample, but do not alter our conclusions that SMGs appear to be consistent with other star-forming galaxies at a given epoch.

The precise values of the blending bias correction factors are likely to be model dependent.
The model used by \citet{2015MNRAS.446.1784C} predicts that close pairs of SMGs are not physically associated and that SMGs singly occupy halos.
There are cases in which blended SMGs are found to be line of sight projections \citep[e.g.][]{2015MNRAS.452.1140Z}, but we note that
\citet{2015ApJ...807..128S} suggest that the detected overabundance of the faint components of multiple SMGs
 in ALMA maps implies that a fraction of the blended SMGs are, on average, physically
associated and are not dominated by line of sight projection effects. Therefore, the model assumed by \citet{2015MNRAS.446.1784C}
may lead to larger blending corrections than is necessary.
Further work will be required to match model SMG clustering properties with observations, which we defer to a future paper. As an extension,
we could input a mock catalogue of galaxies, with
known clustering evolution, into the SCUBA-2 pipeline and tune the input clustering until the clustering of output sources match what is observed. We could therefore predict
more accurately the blending bias from this tuning.
Alternatively, ALMA will soon reveal which SCUBA-2 sources are blends, allowing for a more precise determination of the true galaxy bias.
Based on our current estimates of the likely blending corrections, however, we do not believe these effects will fundamentally alter our conclusions
that SMGs reside in halos comparable to those of star-forming galaxies.

%%%%%%%%%%%%%%%%%%%%%%%%%%%%%%%%%%%%%%%%%%%%%%%%%%%%%%%%%%%%%%%%%%%%%%%%%%%%%%%%%%%%%%%%%%%%%%%%%%%%%%%%%%%%%%%%%%%%%%%%%%%%%%%%%%%%%%%%%%%%%%%%%%%%%%%%%%%%%%%%

\section{CONCLUSIONS}
\label{sec:Conclusion}
We have used a cross-correlation analysis to study the evolution of SMG clustering and halo masses to $z=3.5$. This was made possible by
analysing the largest sample of SMGs in a single field to date, using a combination of $850\,\umu$m flux densities from the SCUBA-2 Cosmology Legacy survey,
radio imaging and optical infra-red
triple colour criteria (OIRTC, developed by \citealt{2016arXiv160102630C}) in the UKIDSS Ultra Deep Survey. We cross-correlate the SMG sample with a much larger K-selected tracer population,
providing significantly improved constraints on halo masses than previously reported. Our main results are summarised as follows.

\begin{enumerate}
  \item We performed a cross-correlation analysis of SMGs in the redshift interval $1<z<3$, the redshift range of focus in previous studies. We find marginally weaker clustering signals
than previously reported, with clustering strength $r_{0}=4.1^{+2.1}_{-2.0}$h$^{-1}$\,Mpc. However, within $1\sigma$ uncertainty, this is consistent with the measurements derived
by earlier smaller-scale studies. We also find that radio-selected SMGs are slightly more strongly clustered, with a clustering strength of $r_{0}=6.8^{+2.7}_{-2.6}$h$^{-1}$\,Mpc.
  \item We analysed the redshift evolution of SMG clustering, deriving halo masses and comparing the measurements with their $UVJ$-selected passive and normal star-forming counterparts.
The clustering strengths of SMGs are consistent with the star-forming population and are lower than that of the passive population at the same redshift.
This finding holds across all redshifts.
We also find tentative evidence of halo downsizing, i.e. SMG activity appears to shift to lower mass dark matter halos as the Universe ages.
The typical halo masses range from M$_{\text{halo}}>10^{13}$\,M$_{\odot}$ at redshifts $z>2.5$, to M$_{\text{halo}}\sim 10^{11}$\,M$_{\odot}$ at redshifts $1<z<2$.
The diminishing clustering signal indicates that SMG activity gradually moves from high-density environments in the early Universe to low-density environments at later epochs.
  \item We tested the previously reported link between local massive elliptical galaxies and SMGs, using a dark matter growth rate formalism.
We find that $z>2$ SMGs are consistent with being the progenitors of massive elliptical galaxies found in present-day galaxy clusters. We note, however, that other star-forming
galaxies at the same epoch show comparable level of clustering. Overall, we find that SMGs exhibit clustering that is consistent with other star-forming galaxies at a given epoch,
with activity shifting from very massive halos at $z>2$ to more modest environments at lower redshift.
\end{enumerate}

%%%%%%%%%%%%%%%%%%%%%%%%%%%%%%%%%%%%%%%%%%%%%%%%%%%%%%%%%%%%%%%%%%%%%%%%%%%%%%%%%%%%%%%%%%%%%%%%%%%%%%%%%%%%%%%%%%%%%%%%%%%%%%%%%%%%%%%%%%%%%%%%%%%%%%%%%%%%%%%%

\section{Acknowledgements}
We would like to thank the anonymous referee for a helpful report, which led to an improvement of the manuscript.
AW acknowledges funding from the STFC. AW wishes to thank the University of Nottingham UDS group for many useful discussions.
IRS and CCC acknowledge support from the STFC (ST/L00075X/1) and the ERC Advanced Grant DUSTYGAL (321334). IRS also acknowledges support from a Royal Society/Wolfson Merit Award.
AM acknowledges funding from the STFC and a European Research Council Consolidator Grant (P.I. R. McLure). V.~W. acknowledges support from the European Research Council
Starting Grant (SEDMorph; P.I. V.~Wild). We extend our gratitude to the staff at UKIRT for their tireless efforts
in ensuring the success of the UDS project. We also wish to recognize and acknowledge the very significant cultural role and reverence that the summit of Mauna Kea has always
had within the indigenous Hawaiian community. We were most fortunate to have the opportunity to conduct observations from this mountain.

\small
\bibliographystyle{mnras}
%\DeclareRobustCommand{\VAN}[3]{#3}
%\bibliographystyle{mn2e}
\bibliography{AWreferences}

\begin{thebibliography}{}
\makeatletter
\relax
\def\mn@urlcharsother{\let\do\@makeother \do\$\do\&\do\#\do\^\do\_\do\%\do\~}
\def\mn@doi{\begingroup\mn@urlcharsother \@ifnextchar [ {\mn@doi@}
  {\mn@doi@[]}}
\def\mn@doi@[#1]#2{\def\@tempa{#1}\ifx\@tempa\@empty \href
  {http://dx.doi.org/#2} {doi:#2}\else \href {http://dx.doi.org/#2} {#1}\fi
  \endgroup}
\def\mn@eprint#1#2{\mn@eprint@#1:#2::\@nil}
\def\mn@eprint@arXiv#1{\href {http://arxiv.org/abs/#1} {{\tt arXiv:#1}}}
\def\mn@eprint@dblp#1{\href {http://dblp.uni-trier.de/rec/bibtex/#1.xml}
  {dblp:#1}}
\def\mn@eprint@#1:#2:#3:#4\@nil{\def\@tempa {#1}\def\@tempb {#2}\def\@tempc
  {#3}\ifx \@tempc \@empty \let \@tempc \@tempb \let \@tempb \@tempa \fi \ifx
  \@tempb \@empty \def\@tempb {arXiv}\fi \@ifundefined
  {mn@eprint@\@tempb}{\@tempb:\@tempc}{\expandafter \expandafter \csname
  mn@eprint@\@tempb\endcsname \expandafter{\@tempc}}}

\bibitem[\protect\citeauthoryear{{Adelberger}, {Steidel}, {Pettini}, {Shapley},
  {Reddy}  \& {Erb}}{{Adelberger} et~al.}{2005}]{2005ApJ...619..697A}
{Adelberger} K.~L.,  {Steidel} C.~C.,  {Pettini} M.,  {Shapley} A.~E.,  {Reddy}
  N.~A.,   {Erb} D.~K.,  2005, \mn@doi [\apj] {10.1086/426580}, \href {http://adsabs.harvard.edu/abs/2005ApJ...619..697A} {619, 697}

\bibitem[\protect\citeauthoryear{{Alexander}, {Bauer}, {Chapman}, {Smail},
  {Blain}, {Brandt}  \& {Ivison}}{{Alexander}
  et~al.}{2005}]{2005ApJ...632..736A}
{Alexander} D.~M.,  {Bauer} F.~E.,  {Chapman} S.~C.,  {Smail} I.,  {Blain}
  A.~W.,  {Brandt} W.~N.,   {Ivison} R.~J.,  2005, \mn@doi [\apj]
  {10.1086/444342}, \href {http://adsabs.harvard.edu/abs/2005ApJ...632..736A} {632, 736}

\bibitem[\protect\citeauthoryear{{Assef} et~al.}{{Assef}
  et~al.}{2011}]{2011ApJ...728...56A}
{Assef} R.~J.,  et~al., 2011, \mn@doi [\apj] {10.1088/0004-637X/728/1/56},
  \href {http://adsabs.harvard.edu/abs/2011ApJ...728...56A} {728, 56}

\bibitem[\protect\citeauthoryear{{Baldry}, {Balogh}, {Bower}, {Glazebrook}  \&
  {Nichol}}{{Baldry} et~al.}{2004}]{2004AIPC..743..106B}
{Baldry} I.~K.,  {Balogh} M.~L.,  {Bower} R.,  {Glazebrook} K.,   {Nichol}
  R.~C.,  2004, in {Allen} R.~E.,  {Nanopoulos} D.~V.,   {Pope} C.~N.,  eds,
  American Institute of Physics Conference Series Vol. 743, The New Cosmology:
  Conference on Strings and Cosmology. pp 106--119 (\mn@eprint {}
  {astro-ph/0410603}), \mn@doi{10.1063/1.1848322}

\bibitem[\protect\citeauthoryear{{Barger}, {Cowie}, {Sanders}, {Fulton},
  {Taniguchi}, {Sato}, {Kawara}  \& {Okuda}}{{Barger}
  et~al.}{1998}]{1998Natur.394..248B}
{Barger} A.~J.,  {Cowie} L.~L.,  {Sanders} D.~B.,  {Fulton} E.,  {Taniguchi}
  Y.,  {Sato} Y.,  {Kawara} K.,   {Okuda} H.,  1998, \mn@doi [\nat]
  {10.1038/28338}, \href {http://adsabs.harvard.edu/abs/1998Natur.394..248B} {394, 248}

\bibitem[\protect\citeauthoryear{{Bell} et~al.}{{Bell}
  et~al.}{2004}]{2004ApJ...608..752B}
{Bell} E.~F.,  et~al., 2004, \mn@doi [\apj] {10.1086/420778}, \href {http://adsabs.harvard.edu/abs/2004ApJ...608..752B} {608, 752}

\bibitem[\protect\citeauthoryear{{Best}, {Kaiser}, {Heckman}  \&
  {Kauffmann}}{{Best} et~al.}{2006}]{2006MNRAS.368L..67B}
{Best} P.~N.,  {Kaiser} C.~R.,  {Heckman} T.~M.,   {Kauffmann} G.,  2006,
  \mn@doi [\mnras] {10.1111/j.1745-3933.2006.00159.x}, \href {http://adsabs.harvard.edu/abs/2006MNRAS.368L..67B} {368, L67}

\bibitem[\protect\citeauthoryear{{Blain}, {Chapman}, {Smail}  \&
  {Ivison}}{{Blain} et~al.}{2004}]{2004ApJ...611..725B}
{Blain} A.~W.,  {Chapman} S.~C.,  {Smail} I.,   {Ivison} R.,  2004, \mn@doi
  [\apj] {10.1086/422353}, \href {http://adsabs.harvard.edu/abs/2004ApJ...611..725B} {611, 725}

\bibitem[\protect\citeauthoryear{{Brammer}, {van Dokkum}  \& {Coppi}}{{Brammer}
  et~al.}{2008}]{2008ApJ...686.1503B}
{Brammer} G.~B.,  {van Dokkum} P.~G.,   {Coppi} P.,  2008, \mn@doi [\apj]
  {10.1086/591786}, \href {http://adsabs.harvard.edu/abs/2008ApJ...686.1503B} {686, 1503}

\bibitem[\protect\citeauthoryear{{Brammer} et~al.}{{Brammer}
  et~al.}{2009}]{2009ApJ...706L.173B}
{Brammer} G.~B.,  et~al., 2009, \mn@doi [\apjl] {10.1088/0004-637X/706/1/L173},
  \href {http://adsabs.harvard.edu/abs/2009ApJ...706L.173B} {706, L173}

\bibitem[\protect\citeauthoryear{{Bruzual} \& {Charlot}}{{Bruzual} \&
  {Charlot}}{2003}]{2003MNRAS.344.1000B}
{Bruzual} G.,  {Charlot} S.,  2003, \mn@doi [\mnras]
  {10.1046/j.1365-8711.2003.06897.x}, \href {http://adsabs.harvard.edu/abs/2003MNRAS.344.1000B} {344, 1000}

\bibitem[\protect\citeauthoryear{{Cen}}{{Cen}}{2011}]{2011ApJ...741...99C}
{Cen} R.,  2011, \mn@doi [\apj] {10.1088/0004-637X/741/2/99}, \href {http://adsabs.harvard.edu/abs/2011ApJ...741...99C} {741, 99}

\bibitem[\protect\citeauthoryear{{Chapman}, {Blain}, {Smail}  \&
  {Ivison}}{{Chapman} et~al.}{2005}]{2005ApJ...622..772C}
{Chapman} S.~C.,  {Blain} A.~W.,  {Smail} I.,   {Ivison} R.~J.,  2005, \mn@doi
  [\apj] {10.1086/428082}, \href {http://adsabs.harvard.edu/abs/2005ApJ...622..772C} {622, 772}

\bibitem[\protect\citeauthoryear{{Chapman}, {Blain}, {Ibata}, {Ivison}, {Smail}
   \& {Morrison}}{{Chapman} et~al.}{2009}]{2009ApJ...691..560C}
{Chapman} S.~C.,  {Blain} A.,  {Ibata} R.,  {Ivison} R.~J.,  {Smail} I.,
  {Morrison} G.,  2009, \mn@doi [\apj] {10.1088/0004-637X/691/1/560}, \href {http://adsabs.harvard.edu/abs/2009ApJ...691..560C} {691, 560}

\bibitem[\protect\citeauthoryear{{Chen} et~al.}{{Chen}
  et~al.}{2016}]{2016arXiv160102630C}
{Chen} C.-C.,  et~al., 2016, preprint, \href {http://adsabs.harvard.edu/abs/2016arXiv160102630C} {} (\mn@eprint {arXiv} {1601.02630})

\bibitem[\protect\citeauthoryear{{Cooray} et~al.}{{Cooray}
  et~al.}{2010}]{Cooray+2010}
{Cooray} A.,  et~al., 2010, \mn@doi [\aap] {10.1051/0004-6361/201014597}, \href {http://adsabs.harvard.edu/abs/2010A%26A...518L..22C} {518, L22}

\bibitem[\protect\citeauthoryear{{Cowley}, {Lacey}, {Baugh}  \&
  {Cole}}{{Cowley} et~al.}{2015a}]{2015arXiv150404516C}
{Cowley} W.~I.,  {Lacey} C.~G.,  {Baugh} C.~M.,   {Cole} S.,  2015a, preprint,
  \href {http://adsabs.harvard.edu/abs/2015arXiv150404516C} {} (\mn@eprint {arXiv} {1504.04516})

\bibitem[\protect\citeauthoryear{{Cowley}, {Lacey}, {Baugh}  \&
  {Cole}}{{Cowley} et~al.}{2015b}]{2015MNRAS.446.1784C}
{Cowley} W.~I.,  {Lacey} C.~G.,  {Baugh} C.~M.,   {Cole} S.,  2015b, \mn@doi
  [\mnras] {10.1093/mnras/stu2179}, \href {http://adsabs.harvard.edu/abs/2015MNRAS.446.1784C} {446, 1784}

\bibitem[\protect\citeauthoryear{{Croton} et~al.}{{Croton}
  et~al.}{2006}]{2006MNRAS.367..864C}
{Croton} D.~J.,  et~al., 2006, \mn@doi [\mnras]
  {10.1111/j.1365-2966.2006.09994.x}, \href {http://adsabs.harvard.edu/abs/2006MNRAS.367..864C} {367, 864}

\bibitem[\protect\citeauthoryear{{Dannerbauer} et~al.,}{{Dannerbauer}
  et~al.}{2014}]{2014A&A...570A..55D}
{Dannerbauer} H.,  et~al., 2014, \mn@doi [\aap] {10.1051/0004-6361/201423771},
  \href {http://adsabs.harvard.edu/abs/2014A%26A...570A..55D} {570, A55}

\bibitem[\protect\citeauthoryear{{Dekel} \& {Birnboim}}{{Dekel} \&
  {Birnboim}}{2006}]{2006MNRAS.368....2D}
{Dekel} A.,  {Birnboim} Y.,  2006, \mn@doi [\mnras]
  {10.1111/j.1365-2966.2006.10145.x}, \href {http://adsabs.harvard.edu/abs/2006MNRAS.368....2D} {368, 2}

\bibitem[\protect\citeauthoryear{{Dekel}, {Sari}  \& {Ceverino}}{{Dekel}
  et~al.}{2009}]{2009ApJ...703..785D}
{Dekel} A.,  {Sari} R.,   {Ceverino} D.,  2009, \mn@doi [\apj]
  {10.1088/0004-637X/703/1/785}, \href {http://adsabs.harvard.edu/abs/2009ApJ...703..785D} {703, 785}

\bibitem[\protect\citeauthoryear{{Di Matteo}, {Springel}  \& {Hernquist}}{{Di
  Matteo} et~al.}{2005}]{2005Natur.433..604D}
{Di Matteo} T.,  {Springel} V.,   {Hernquist} L.,  2005, \mn@doi [\nat]
  {10.1038/nature03335}, \href {http://adsabs.harvard.edu/abs/2005Natur.433..604D} {433, 604}

\bibitem[\protect\citeauthoryear{{Eales}, {Lilly}, {Gear}, {Dunne}, {Bond},
  {Hammer}, {Le F{\`e}vre}  \& {Crampton}}{{Eales}
  et~al.}{1999}]{1999ApJ...515..518E}
{Eales} S.,  {Lilly} S.,  {Gear} W.,  {Dunne} L.,  {Bond} J.~R.,  {Hammer} F.,
  {Le F{\`e}vre} O.,   {Crampton} D.,  1999, \mn@doi [\apj] {10.1086/307069},
  \href {http://adsabs.harvard.edu/abs/1999ApJ...515..518E} {515, 518}

\bibitem[\protect\citeauthoryear{{Estrada}, {Sefusatti}  \&
  {Frieman}}{{Estrada} et~al.}{2009}]{2009ApJ...692..265E}
{Estrada} J.,  {Sefusatti} E.,   {Frieman} J.~A.,  2009, \mn@doi [\apj]
  {10.1088/0004-637X/692/1/265}, \href {http://adsabs.harvard.edu/abs/2009ApJ...692..265E} {692, 265}

\bibitem[\protect\citeauthoryear{{Fakhouri}, {Ma}  \&
  {Boylan-Kolchin}}{{Fakhouri} et~al.}{2010}]{2010MNRAS.406.2267F}
{Fakhouri} O.,  {Ma} C.-P.,   {Boylan-Kolchin} M.,  2010, \mn@doi [\mnras]
  {10.1111/j.1365-2966.2010.16859.x}, \href {http://adsabs.harvard.edu/abs/2010MNRAS.406.2267F} {406, 2267}

\bibitem[\protect\citeauthoryear{{Farrah} et~al.,}{{Farrah}
  et~al.}{2006}]{2006ApJ...641L..17F}
{Farrah} D.,  et~al., 2006, \mn@doi [\apjl] {10.1086/503769}, \href {http://adsabs.harvard.edu/abs/2006ApJ...641L..17F} {641, L17}

\bibitem[\protect\citeauthoryear{{Furusawa} et~al.}{{Furusawa}
  et~al.}{2008}]{2008ApJS..176....1F}
{Furusawa} H.,  et~al., 2008, \mn@doi [\apjs] {10.1086/527321}, \href {http://adsabs.harvard.edu/abs/2008ApJS..176....1F} {176, 1}

\bibitem[\protect\citeauthoryear{{Furusawa}, {Sekiguchi}, {Takata}, {Furusawa},
  {Shimasaku}, {Simpson}  \& {Akiyama}}{{Furusawa}
  et~al.}{2011}]{2011ApJ...727..111F}
{Furusawa} J.,  {Sekiguchi} K.,  {Takata} T.,  {Furusawa} H.,  {Shimasaku} K.,
  {Simpson} C.,   {Akiyama} M.,  2011, \mn@doi [\apj]
  {10.1088/0004-637X/727/2/111}, \href {http://adsabs.harvard.edu/abs/2011ApJ...727..111F} {727, 111}

\bibitem[\protect\citeauthoryear{{Geach} et~al.}{{Geach}
  et~al.}{2013}]{2013MNRAS.432...53G}
{Geach} J.~E.,  et~al., 2013, \mn@doi [\mnras] {10.1093/mnras/stt352}, \href {http://adsabs.harvard.edu/abs/2013MNRAS.432...53G} {432, 53}

\bibitem[\protect\citeauthoryear{{Hainline}, {Blain}, {Smail}, {Alexander},
  {Armus}, {Chapman}  \& {Ivison}}{{Hainline}
  et~al.}{2011}]{2011ApJ...740...96H}
{Hainline} L.~J.,  {Blain} A.~W.,  {Smail} I.,  {Alexander} D.~M.,  {Armus} L.,
   {Chapman} S.~C.,   {Ivison} R.~J.,  2011, \mn@doi [\apj]
  {10.1088/0004-637X/740/2/96}, \href {http://adsabs.harvard.edu/abs/2011ApJ...740...96H} {740, 96}

\bibitem[\protect\citeauthoryear{{Hartley} et~al.}{{Hartley}
  et~al.}{2010}]{2010MNRAS.407.1212H}
{Hartley} W.~G.,  et~al., 2010, \mn@doi [\mnras]
  {10.1111/j.1365-2966.2010.16972.x}, \href {http://adsabs.harvard.edu/abs/2010MNRAS.407.1212H} {407, 1212}

\bibitem[\protect\citeauthoryear{{Hartley} et~al.}{{Hartley}
  et~al.}{2013}]{2013MNRAS.431.3045H}
{Hartley} W.~G.,  et~al., 2013, \mn@doi [\mnras] {10.1093/mnras/stt383}, \href {http://adsabs.harvard.edu/abs/2013MNRAS.431.3045H} {431, 3045}

\bibitem[\protect\citeauthoryear{{Hayward}, {Behroozi}, {Somerville},
  {Primack}, {Moreno}  \& {Wechsler}}{{Hayward}
  et~al.}{2013}]{2013MNRAS.434.2572H}
{Hayward} C.~C.,  {Behroozi} P.~S.,  {Somerville} R.~S.,  {Primack} J.~R.,
  {Moreno} J.,   {Wechsler} R.~H.,  2013, \mn@doi [\mnras]
  {10.1093/mnras/stt1202}, \href {http://adsabs.harvard.edu/abs/2013MNRAS.434.2572H} {434, 2572}

\bibitem[\protect\citeauthoryear{{Hickox} et~al.}{{Hickox}
  et~al.}{2012}]{2012MNRAS.421..284H}
{Hickox} R.~C.,  et~al., 2012, \mn@doi [\mnras]
  {10.1111/j.1365-2966.2011.20303.x}, \href {http://adsabs.harvard.edu/abs/2012MNRAS.421..284H} {421, 284}

\bibitem[\protect\citeauthoryear{{Hodge} et~al.}{{Hodge}
  et~al.}{2013}]{2013ApJ...768...91H}
{Hodge} J.~A.,  et~al., 2013, \mn@doi [\apj] {10.1088/0004-637X/768/1/91},
  \href {http://adsabs.harvard.edu/abs/2013ApJ...768...91H} {768, 91}

\bibitem[\protect\citeauthoryear{{Holland} et~al.}{{Holland}
  et~al.}{2013}]{2013MNRAS.430.2513H}
{Holland} W.~S.,  et~al., 2013, \mn@doi [\mnras] {10.1093/mnras/sts612}, \href {http://adsabs.harvard.edu/abs/2013MNRAS.430.2513H} {430, 2513}

\bibitem[\protect\citeauthoryear{{Hopkins} et~al.}{{Hopkins}
  et~al.}{2005}]{2005ApJ...630..705H}
{Hopkins} P.~F.,  et~al., 2005, \mn@doi [\apj] {10.1086/432438}, \href {http://adsabs.harvard.edu/abs/2005ApJ...630..705H} {630, 705}

\bibitem[\protect\citeauthoryear{{Hopkins}, {Hernquist}, {Cox}, {Di Matteo},
  {Robertson}  \& {Springel}}{{Hopkins} et~al.}{2006}]{2006ApJS..163....1H}
{Hopkins} P.~F.,  {Hernquist} L.,  {Cox} T.~J.,  {Di Matteo} T.,  {Robertson}
  B.,   {Springel} V.,  2006, \mn@doi [\apjs] {10.1086/499298}, \href {http://adsabs.harvard.edu/abs/2006ApJS..163....1H} {163, 1}

\bibitem[\protect\citeauthoryear{{Hughes} et~al.}{{Hughes}
  et~al.}{1998}]{1998Natur.394..241H}
{Hughes} D.~H.,  et~al., 1998, \mn@doi [\nat] {10.1038/28328}, \href {http://adsabs.harvard.edu/abs/1998Natur.394..241H} {394, 241}

\bibitem[\protect\citeauthoryear{{Ivison} et~al.}{{Ivison}
  et~al.}{2002}]{2002MNRAS.337....1I}
{Ivison} R.~J.,  et~al., 2002, \mn@doi [\mnras]
  {10.1046/j.1365-8711.2002.05900.x}, \href {http://adsabs.harvard.edu/abs/2002MNRAS.337....1I} {337, 1}

\bibitem[\protect\citeauthoryear{{Ivison} et~al.}{{Ivison}
  et~al.}{2005}]{2005MNRAS.364.1025I}
{Ivison} R.~J.,  et~al., 2005, \mn@doi [\mnras]
  {10.1111/j.1365-2966.2005.09639.x}, \href {http://adsabs.harvard.edu/abs/2005MNRAS.364.1025I} {364, 1025}

\bibitem[\protect\citeauthoryear{{Ivison}, {Papadopoulos}, {Smail}, {Greve},
  {Thomson}, {Xilouris}  \& {Chapman}}{{Ivison}
  et~al.}{2011}]{2011MNRAS.412.1913I}
{Ivison} R.~J.,  {Papadopoulos} P.~P.,  {Smail} I.,  {Greve} T.~R.,  {Thomson}
  A.~P.,  {Xilouris} E.~M.,   {Chapman} S.~C.,  2011, \mn@doi [\mnras]
  {10.1111/j.1365-2966.2010.18028.x}, \href {http://adsabs.harvard.edu/abs/2011MNRAS.412.1913I} {412, 1913}

\bibitem[\protect\citeauthoryear{{Jullo} et~al.}{{Jullo}
  et~al.}{2012}]{2012ApJ...750...37J}
{Jullo} E.,  et~al., 2012, \mn@doi [\apj] {10.1088/0004-637X/750/1/37}, \href {http://adsabs.harvard.edu/abs/2012ApJ...750...37J} {750, 37}

\bibitem[\protect\citeauthoryear{{Karim} et~al.}{{Karim}
  et~al.}{2013}]{2013MNRAS.432....2K}
{Karim} A.,  et~al., 2013, \mn@doi [\mnras] {10.1093/mnras/stt196}, \href {http://adsabs.harvard.edu/abs/2013MNRAS.432....2K} {432, 2}

\bibitem[\protect\citeauthoryear{{Kennicutt}}{{Kennicutt}}{1998}]{1998ARA&A..3%
6..189K}
{Kennicutt} Jr. R.~C.,  1998, \mn@doi [\araa] {10.1146/annurev.astro.36.1.189},
  \href {http://adsabs.harvard.edu/abs/1998ARA%26A..36..189K} {36, 189}

\bibitem[\protect\citeauthoryear{{Lacey} et~al.}{{Lacey}
  et~al.}{2015}]{2015arXiv150908473L}
{Lacey} C.~G.,  et~al., 2015, preprint, \href {http://adsabs.harvard.edu/abs/2015arXiv150908473L} {} (\mn@eprint {arXiv} {1509.08473})

\bibitem[\protect\citeauthoryear{{Landy} \& {Szalay}}{{Landy} \&
  {Szalay}}{1993}]{1993ApJ...412...64L}
{Landy} S.~D.,  {Szalay} A.~S.,  1993, \mn@doi [\apj] {10.1086/172900}, \href {http://adsabs.harvard.edu/abs/1993ApJ...412...64L} {412, 64}

\bibitem[\protect\citeauthoryear{{Lin} et~al.}{{Lin}
  et~al.}{2016}]{2016ApJ...817...97L}
{Lin} L.,  et~al., 2016, \mn@doi [\apj] {10.3847/0004-637X/817/2/97}, \href {http://adsabs.harvard.edu/abs/2016ApJ...817...97L} {817, 97}

\bibitem[\protect\citeauthoryear{{Maddox} et~al.}{{Maddox}
  et~al.}{2010}]{2010A&A...518L..11M}
{Maddox} S.~J.,  et~al., 2010, \mn@doi [\aap] {10.1051/0004-6361/201014663},
  \href {http://adsabs.harvard.edu/abs/2010A%26A...518L..11M} {518, L11}

\bibitem[\protect\citeauthoryear{{Magliocchetti} \& {Porciani}}{{Magliocchetti}
  \& {Porciani}}{2003}]{2003MNRAS.346..186M}
{Magliocchetti} M.,  {Porciani} C.,  2003, \mn@doi [\mnras]
  {10.1046/j.1365-2966.2003.07094.x}, \href {http://adsabs.harvard.edu/abs/2003MNRAS.346..186M} {346, 186}

\bibitem[\protect\citeauthoryear{{Magliocchetti} et~al.,}{{Magliocchetti}
  et~al.}{2013}]{2013MNRAS.433..127M}
{Magliocchetti} M.,  et~al., 2013, \mn@doi [\mnras] {10.1093/mnras/stt708},
  \href {http://adsabs.harvard.edu/abs/2013MNRAS.433..127M} {433, 127}

\bibitem[\protect\citeauthoryear{{Martig}, {Bournaud}, {Teyssier}  \&
  {Dekel}}{{Martig} et~al.}{2009}]{2009ApJ...707..250M}
{Martig} M.,  {Bournaud} F.,  {Teyssier} R.,   {Dekel} A.,  2009, \mn@doi
  [\apj] {10.1088/0004-637X/707/1/250}, \href {http://adsabs.harvard.edu/abs/2009ApJ...707..250M} {707, 250}

\bibitem[\protect\citeauthoryear{{McCracken} et~al.}{{McCracken}
  et~al.}{2015}]{2015MNRAS.449..901M}
{McCracken} H.~J.,  et~al., 2015, \mn@doi [\mnras] {10.1093/mnras/stv305},
  \href {http://adsabs.harvard.edu/abs/2015MNRAS.449..901M} {449, 901}

\bibitem[\protect\citeauthoryear{{Meneux} et~al.}{{Meneux}
  et~al.}{2006}]{2006A&A...452..387M}
{Meneux} B.,  et~al., 2006, \mn@doi [\aap] {10.1051/0004-6361:20054571}, \href {http://adsabs.harvard.edu/abs/2006A%26A...452..387M} {452, 387}

\bibitem[\protect\citeauthoryear{{Miller}, {Hayward}, {Chapman}  \&
  {Behroozi}}{{Miller} et~al.}{2015}]{2015MNRAS.452..878M}
{Miller} T.~B.,  {Hayward} C.~C.,  {Chapman} S.~C.,   {Behroozi} P.~S.,  2015,
  \mn@doi [\mnras] {10.1093/mnras/stv1267}, \href {http://adsabs.harvard.edu/abs/2015MNRAS.452..878M} {452, 878}

\bibitem[\protect\citeauthoryear{{Mitchell-Wynne} et~al.}{{Mitchell-Wynne}
  et~al.}{2012}]{2012ApJ...753...23M}
{Mitchell-Wynne} K.,  et~al., 2012, \mn@doi [\apj]
  {10.1088/0004-637X/753/1/23}, \href {http://adsabs.harvard.edu/abs/2012ApJ...753...23M} {753, 23}

\bibitem[\protect\citeauthoryear{{Mo} \& {White}}{{Mo} \&
  {White}}{1996}]{1996MNRAS.282..347M}
{Mo} H.~J.,  {White} S.~D.~M.,  1996, \mn@doi [\mnras]
  {10.1093/mnras/282.2.347}, \href {http://adsabs.harvard.edu/abs/1996MNRAS.282..347M} {282, 347}

\bibitem[\protect\citeauthoryear{{Mo} \& {White}}{{Mo} \&
  {White}}{2002}]{2002MNRAS.336..112M}
{Mo} H.~J.,  {White} S.~D.~M.,  2002, \mn@doi [\mnras]
  {10.1046/j.1365-8711.2002.05723.x}, \href {http://adsabs.harvard.edu/abs/2002MNRAS.336..112M} {336, 112}

\bibitem[\protect\citeauthoryear{{Mortlock} et~al.}{{Mortlock}
  et~al.}{2013}]{2013MNRAS.433.1185M}
{Mortlock} A.,  et~al., 2013, \mn@doi [\mnras] {10.1093/mnras/stt793}, \href {http://adsabs.harvard.edu/abs/2013MNRAS.433.1185M} {433, 1185}

\bibitem[\protect\citeauthoryear{{Mortlock} et~al.}{{Mortlock}
  et~al.}{2015}]{2015MNRAS.447....2M}
{Mortlock} A.,  et~al., 2015, \mn@doi [\mnras] {10.1093/mnras/stu2403}, \href {http://adsabs.harvard.edu/abs/2015MNRAS.447....2M} {447, 2}

\bibitem[\protect\citeauthoryear{{Norberg} et~al.}{{Norberg}
  et~al.}{2002}]{2002MNRAS.332..827N}
{Norberg} P.,  et~al., 2002, \mn@doi [\mnras]
  {10.1046/j.1365-8711.2002.05348.x}, \href {http://adsabs.harvard.edu/abs/2002MNRAS.332..827N} {332, 827}

\bibitem[\protect\citeauthoryear{{Peebles}}{{Peebles}}{1980}]{1980lssu.book...%
..P}
{Peebles} P.~J.~E.,  1980, {The large-scale structure of the universe}

\bibitem[\protect\citeauthoryear{{Pope} et~al.}{{Pope}
  et~al.}{2006}]{2006MNRAS.370.1185P}
{Pope} A.,  et~al., 2006, \mn@doi [\mnras] {10.1111/j.1365-2966.2006.10575.x},
  \href {http://adsabs.harvard.edu/abs/2006MNRAS.370.1185P} {370, 1185}

\bibitem[\protect\citeauthoryear{{Pozzetti} et~al.}{{Pozzetti}
  et~al.}{2010}]{2010A&A...523A..13P}
{Pozzetti} L.,  et~al., 2010, \mn@doi [\aap] {10.1051/0004-6361/200913020},
  \href {http://adsabs.harvard.edu/abs/2010A%26A...523A..13P} {523, A13}

\bibitem[\protect\citeauthoryear{{Prevot}, {Lequeux}, {Prevot}, {Maurice}  \&
  {Rocca-Volmerange}}{{Prevot} et~al.}{1984}]{1984A&A...132..389P}
{Prevot} M.~L.,  {Lequeux} J.,  {Prevot} L.,  {Maurice} E.,
  {Rocca-Volmerange} B.,  1984, \aap, \href {http://adsabs.harvard.edu/abs/1984A%26A...132..389P} {132, 389}

\bibitem[\protect\citeauthoryear{{Richards} et~al.}{{Richards}
  et~al.}{2006}]{2006AJ....131.2766R}
{Richards} G.~T.,  et~al., 2006, \mn@doi [\aj] {10.1086/503559}, \href {http://adsabs.harvard.edu/abs/2006AJ....131.2766R} {131, 2766}

\bibitem[\protect\citeauthoryear{{Roche} \& {Eales}}{{Roche} \&
  {Eales}}{1999}]{1999MNRAS.307..703R}
{Roche} N.,  {Eales} S.~A.,  1999, \mn@doi [\mnras]
  {10.1046/j.1365-8711.1999.02652.x}, \href {http://adsabs.harvard.edu/abs/1999MNRAS.307..703R} {307, 703}

\bibitem[\protect\citeauthoryear{{Ross} \& {Brunner}}{{Ross} \&
  {Brunner}}{2009}]{2009MNRAS.399..878R}
{Ross} A.~J.,  {Brunner} R.~J.,  2009, \mn@doi [\mnras]
  {10.1111/j.1365-2966.2009.15318.x}, \href {http://adsabs.harvard.edu/abs/2009MNRAS.399..878R} {399, 878}

\bibitem[\protect\citeauthoryear{{Silk} \& {Rees}}{{Silk} \&
  {Rees}}{1998}]{1998A&A...331L...1S}
{Silk} J.,  {Rees} M.~J.,  1998, \aap, \href {http://adsabs.harvard.edu/abs/1998A%26A...331L...1S} {331, L1}

\bibitem[\protect\citeauthoryear{{Simpson} et~al.}{{Simpson}
  et~al.}{2006}]{2006MNRAS.372..741S}
{Simpson} C.,  et~al., 2006, \mn@doi [\mnras]
  {10.1111/j.1365-2966.2006.10907.x}, \href {http://adsabs.harvard.edu/abs/2006MNRAS.372..741S} {372, 741}

\bibitem[\protect\citeauthoryear{{Simpson} et~al.}{{Simpson}
  et~al.}{2014}]{2014ApJ...788..125S}
{Simpson} J.~M.,  et~al., 2014, \mn@doi [\apj] {10.1088/0004-637X/788/2/125},
  \href {http://adsabs.harvard.edu/abs/2014ApJ...788..125S} {788, 125}

\bibitem[\protect\citeauthoryear{{Simpson} et~al.}{{Simpson}
  et~al.}{2015a}]{2015ApJ...799...81S}
{Simpson} J.~M.,  et~al., 2015a, \mn@doi [\apj] {10.1088/0004-637X/799/1/81},
  \href {http://adsabs.harvard.edu/abs/2015ApJ...799...81S} {799, 81}

\bibitem[\protect\citeauthoryear{{Simpson} et~al.}{{Simpson}
  et~al.}{2015b}]{2015ApJ...807..128S}
{Simpson} J.~M.,  et~al., 2015b, \mn@doi [\apj] {10.1088/0004-637X/807/2/128},
  \href {http://adsabs.harvard.edu/abs/2015ApJ...807..128S} {807, 128}

\bibitem[\protect\citeauthoryear{{Smail}, {Ivison}  \& {Blain}}{{Smail}
  et~al.}{1997}]{1997ApJ...490L...5S}
{Smail} I.,  {Ivison} R.~J.,   {Blain} A.~W.,  1997, \mn@doi [\apjl]
  {10.1086/311017}, \href {http://adsabs.harvard.edu/abs/1997ApJ...490L...5S}
  {490, L5}

\bibitem[\protect\citeauthoryear{{Smail} et~al.}{{Smail}
  et~al.}{2014}]{2014ApJ...782...19S}
{Smail} I.,  et~al., 2014, \mn@doi [\apj] {10.1088/0004-637X/782/1/19}, \href {http://adsabs.harvard.edu/abs/2014ApJ...782...19S} {782, 19}

\bibitem[\protect\citeauthoryear{{Smith} et~al.}{{Smith}
  et~al.}{2003}]{2003MNRAS.341.1311S}
{Smith} R.~E.,  et~al., 2003, \mn@doi [\mnras]
  {10.1046/j.1365-8711.2003.06503.x}, \href {http://adsabs.harvard.edu/abs/2003MNRAS.341.1311S} {341, 1311}

\bibitem[\protect\citeauthoryear{{Somerville}, {Hopkins}, {Cox}, {Robertson}
  \& {Hernquist}}{{Somerville} et~al.}{2008}]{2008MNRAS.391..481S}
{Somerville} R.~S.,  {Hopkins} P.~F.,  {Cox} T.~J.,  {Robertson} B.~E.,
  {Hernquist} L.,  2008, \mn@doi [\mnras] {10.1111/j.1365-2966.2008.13805.x},
  \href {http://adsabs.harvard.edu/abs/2008MNRAS.391..481S} {391, 481}

\bibitem[\protect\citeauthoryear{{Steidel}, {Adelberger}, {Dickinson},
  {Giavalisco}, {Pettini}  \& {Kellogg}}{{Steidel}
  et~al.}{1998}]{1998ApJ...492..428S}
{Steidel} C.~C.,  {Adelberger} K.~L.,  {Dickinson} M.,  {Giavalisco} M.,
  {Pettini} M.,   {Kellogg} M.,  1998, \mn@doi [\apj] {10.1086/305073}, \href {http://adsabs.harvard.edu/abs/1998ApJ...492..428S} {492, 428}

\bibitem[\protect\citeauthoryear{{Strateva} et~al.}{{Strateva}
  et~al.}{2001}]{2001AJ....122.1861S}
{Strateva} I.,  et~al., 2001, \mn@doi [\aj] {10.1086/323301}, \href {http://adsabs.harvard.edu/abs/2001AJ....122.1861S} {122, 1861}

\bibitem[\protect\citeauthoryear{{Swinbank}, {Chapman}, {Smail}, {Lindner},
  {Borys}, {Blain}, {Ivison}  \& {Lewis}}{{Swinbank}
  et~al.}{2006}]{2006MNRAS.371..465S}
{Swinbank} A.~M.,  {Chapman} S.~C.,  {Smail} I.,  {Lindner} C.,  {Borys} C.,
  {Blain} A.~W.,  {Ivison} R.~J.,   {Lewis} G.~F.,  2006, \mn@doi [\mnras]
  {10.1111/j.1365-2966.2006.10673.x}, \href {http://adsabs.harvard.edu/abs/2006MNRAS.371..465S} {371, 465}

\bibitem[\protect\citeauthoryear{{Tacconi} et~al.}{{Tacconi}
  et~al.}{2006}]{2006ApJ...640..228T}
{Tacconi} L.~J.,  et~al., 2006, \mn@doi [\apj] {10.1086/499933}, \href {http://adsabs.harvard.edu/abs/2006ApJ...640..228T} {640, 228}

\bibitem[\protect\citeauthoryear{{Tacconi} et~al.}{{Tacconi}
  et~al.}{2008}]{2008ApJ...680..246T}
{Tacconi} L.~J.,  et~al., 2008, \mn@doi [\apj] {10.1086/587168}, \href {http://adsabs.harvard.edu/abs/2008ApJ...680..246T} {680, 246}

\bibitem[\protect\citeauthoryear{{Targett}, {Dunlop}, {McLure}, {Best},
  {Cirasuolo}  \& {Almaini}}{{Targett} et~al.}{2011}]{2011MNRAS.412..295T}
{Targett} T.~A.,  {Dunlop} J.~S.,  {McLure} R.~J.,  {Best} P.~N.,  {Cirasuolo}
  M.,   {Almaini} O.,  2011, \mn@doi [\mnras]
  {10.1111/j.1365-2966.2010.17905.x}, \href {http://adsabs.harvard.edu/abs/2011MNRAS.412..295T} {412, 295}

\bibitem[\protect\citeauthoryear{{Thomas}, {Maraston}, {Schawinski}, {Sarzi}
  \& {Silk}}{{Thomas} et~al.}{2010}]{2010MNRAS.404.1775T}
{Thomas} D.,  {Maraston} C.,  {Schawinski} K.,  {Sarzi} M.,   {Silk} J.,  2010,
  \mn@doi [\mnras] {10.1111/j.1365-2966.2010.16427.x}, \href {http://adsabs.harvard.edu/abs/2010MNRAS.404.1775T} {404, 1775}

\bibitem[\protect\citeauthoryear{{Toomre}}{{Toomre}}{1977}]{1977egsp.conf..401%
T}
{Toomre} A.,  1977, in {Tinsley} B.~M.,  {Larson} D.~Campbell R.~B.~G.,  eds,
  Evolution of Galaxies and Stellar Populations. p.~401

\bibitem[\protect\citeauthoryear{{Trayford}, {Theuns}, {Bower}, {Crain},
  {Lagos}, {Schaller}  \& {Schaye}}{{Trayford}
  et~al.}{2016}]{2016arXiv160107907T}
{Trayford} J.~W.,  {Theuns} T.,  {Bower} R.~G.,  {Crain} R.~A.,  {Lagos}
  C.~d.~P.,  {Schaller} M.,   {Schaye} J.,  2016, preprint, \href {http://adsabs.harvard.edu/abs/2016arXiv160107907T} {} (\mn@eprint {arXiv}
  {1601.07907})

\bibitem[\protect\citeauthoryear{{Ueda} et~al.}{{Ueda}
  et~al.}{2008}]{2008ApJS..179..124U}
{Ueda} Y.,  et~al., 2008, \mn@doi [\apjs] {10.1086/591083}, \href {http://adsabs.harvard.edu/abs/2008ApJS..179..124U} {179, 124}

\bibitem[\protect\citeauthoryear{{Umehata} et~al.}{{Umehata}
  et~al.}{2015}]{2015ApJ...815L...8U}
{Umehata} H.,  et~al., 2015, \mn@doi [\apjl] {10.1088/2041-8205/815/1/L8},
  \href {http://adsabs.harvard.edu/abs/2015ApJ...815L...8U} {815, L8}

\bibitem[\protect\citeauthoryear{{van}~Kampen et~al.}{{van}~Kampen
  et~al.}{2012}]{2012MNRAS.426.3455V}
{van}~Kampen E.,  et~al., 2012, \mn@doi [\mnras]
  {10.1111/j.1365-2966.2012.21949.x}, \href {http://adsabs.harvard.edu/abs/2012MNRAS.426.3455V} {426, 3455}

\bibitem[\protect\citeauthoryear{{Wake} et~al.}{{Wake}
  et~al.}{2006}]{2006MNRAS.372..537W}
{Wake} D.~A.,  et~al., 2006, \mn@doi [\mnras]
  {10.1111/j.1365-2966.2006.10831.x}, \href {http://adsabs.harvard.edu/abs/2006MNRAS.372..537W} {372, 537}

\bibitem[\protect\citeauthoryear{{Wake} et~al.}{{Wake}
  et~al.}{2011}]{2011ApJ...728...46W}
{Wake} D.~A.,  et~al., 2011, \mn@doi [\apj] {10.1088/0004-637X/728/1/46}, \href {http://adsabs.harvard.edu/abs/2011ApJ...728...46W} {728, 46}

\bibitem[\protect\citeauthoryear{{Webb} et~al.}{{Webb}
  et~al.}{2003}]{2003ApJ...582....6W}
{Webb} T.~M.,  et~al., 2003, \mn@doi [\apj] {10.1086/344608}, \href {http://adsabs.harvard.edu/abs/2003ApJ...582....6W} {582, 6}

\bibitem[\protect\citeauthoryear{{Wei{\ss}} et~al.}{{Wei{\ss}}
  et~al.}{2009}]{2009ApJ...707.1201W}
{Wei{\ss}} A.,  et~al., 2009, \mn@doi [\apj] {10.1088/0004-637X/707/2/1201},
  \href {http://adsabs.harvard.edu/abs/2009ApJ...707.1201W} {707, 1201}

\bibitem[\protect\citeauthoryear{{Wild} et~al.}{{Wild}
  et~al.}{2014}]{2014MNRAS.440.1880W}
{Wild} V.,  et~al., 2014, \mn@doi [\mnras] {10.1093/mnras/stu212}, \href {http://adsabs.harvard.edu/abs/2014MNRAS.440.1880W} {440, 1880}

\bibitem[\protect\citeauthoryear{{Williams}, {Quadri}, {Franx}, {van Dokkum}
  \& {Labb{\'e}}}{{Williams} et~al.}{2009}]{2009ApJ...691.1879W}
{Williams} R.~J.,  {Quadri} R.~F.,  {Franx} M.,  {van Dokkum} P.,   {Labb{\'e}}
  I.,  2009, \mn@doi [\apj] {10.1088/0004-637X/691/2/1879}, \href {http://adsabs.harvard.edu/abs/2009ApJ...691.1879W} {691, 1879}

\bibitem[\protect\citeauthoryear{{Williams} et~al.}{{Williams}
  et~al.}{2011}]{2011ApJ...733...92W}
{Williams} C.~C.,  et~al., 2011, \mn@doi [\apj] {10.1088/0004-637X/733/2/92},
  \href {http://adsabs.harvard.edu/abs/2011ApJ...733...92W} {733, 92}

\bibitem[\protect\citeauthoryear{{Yun} et~al.}{{Yun}
  et~al.}{2012}]{2012MNRAS.420..957Y}
{Yun} M.~S.,  et~al., 2012, \mn@doi [\mnras]
  {10.1111/j.1365-2966.2011.19898.x}, \href {http://adsabs.harvard.edu/abs/2012MNRAS.420..957Y} {420, 957}

\bibitem[\protect\citeauthoryear{{Zavala} et~al.}{{Zavala}
  et~al.}{2015}]{2015MNRAS.452.1140Z}
{Zavala} J.~A.,  et~al., 2015, \mn@doi [\mnras] {10.1093/mnras/stv1351}, \href {http://adsabs.harvard.edu/abs/2015MNRAS.452.1140Z} {452, 1140}

\bibitem[\protect\citeauthoryear{{Zehavi} et~al.}{{Zehavi}
  et~al.}{2011}]{2011ApJ...736...59Z}
{Zehavi} I.,  et~al., 2011, \mn@doi [\apj] {10.1088/0004-637X/736/1/59}, \href {http://adsabs.harvard.edu/abs/2011ApJ...736...59Z} {736, 59}


\makeatother
\end{thebibliography}

%\begin{thebibliography}{99}
%\bibitem[\protect\citeauthoryear

%\end{thebibliography}

\bsp

\label{lastpage}

\end{document}